%% file: main.tex
\documentclass[10pt,twocolumn,letterpaper]{article}

\usepackage{iccv} 

\usepackage{times}
\usepackage{epsfig}
\usepackage{graphicx}
\usepackage{amsmath}
\usepackage{amssymb}
\usepackage{multirow}
\usepackage{adjustbox}
\usepackage{fontawesome}
\usepackage{xcolor,colortbl}
\definecolor{light}{gray}{0.95}

\definecolor{iccvblue}{rgb}{0.21,0.49,0.74}
\usepackage[pagebackref,breaklinks,colorlinks,allcolors=iccvblue]{hyperref}

\def\ourMethod{GeometryCrafter}

\title{{\ourMethod}: Consistent Geometry Estimation for Open-world Videos \\ with Diffusion Priors}

\author{Tian-Xing Xu$^{1}$\quad Xiangjun Gao$^{3}$\quad Wenbo Hu$^{2 \, \dagger}$ \quad Xiaoyu Li$^{2}$ \quad Song-Hai Zhang$^{1 \, \dagger}$ \quad Ying Shan$^{2}$\\
$^{1}$ Tsinghua University \quad $^{2}$ ARC Lab, Tencent PCG \quad $^{3}$ HKUST \\
{\tt\small Project Page: \url{https://geometrycrafter.github.io}}
}

\newcommand{\first}[1]{\textbf{#1}}
\newcommand{\second}[1]{\underline{#1}}
\newcommand{\notzs}[1]{\textcolor[rgb]{0.5, 0.5, 0.5}{#1$^*$}}
\newcommand{\boldstartspace}[1]{\medskip\noindent\textbf{#1}}

\begin{document}

\twocolumn[{%
\renewcommand\twocolumn[1][]{#1}%
\maketitle
\vspace{-2em}
\includegraphics[width=.97\linewidth]{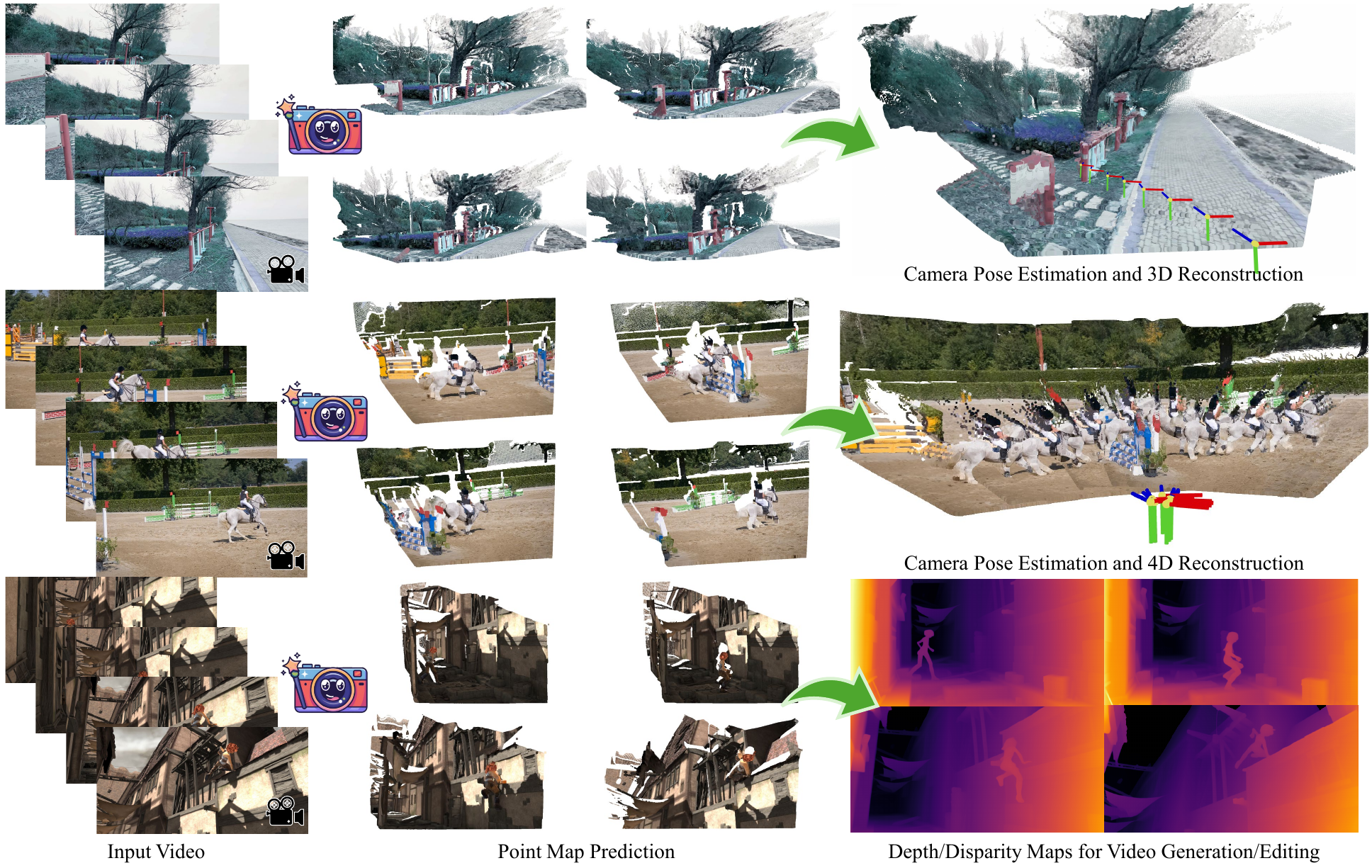}
\vspace{-0.8em}
\captionof{figure}{
    We present {\ourMethod}, a novel approach that estimates temporally consistent, high-quality point maps from open-world videos, facilitating downstream applications such as 3D/4D reconstruction and depth-based video editing or generation.
    }
\vspace{1em}
\label{fig:teaser}
}]

\renewcommand{\thefootnote}{}
\footnotetext[0]{ $^\dag$Corresponding authors.}

\input{sections/abstract}    
\input{sections/introduction}    
\input{sections/related_works}    
\input{sections/method}

\input{sections/experiments}

\input{sections/application}
\input{sections/conclusion}

\section*{Acknowledgement}

Tian-Xing Xu completed this work during his internship at Tencent ARC Lab. The project was supported by the Tsinghua-Tencent Joint Laboratory for Internet Innovation Technology.


{
    \small
    \bibliographystyle{ieeenat_fullname}
    \bibliography{main}
}

\clearpage
\setcounter{page}{1}
\setcounter{section}{0}
\setcounter{table}{0}
\setcounter{figure}{0}
\maketitlesupplementary

\input{supps/dataset} 
\input{supps/losses}
\input{supps/implementation}
\input{supps/camera_pose}
\input{supps/limitations}

\input{supps/results}

\end{document}

%% file: sections/abstract.tex
\begin{abstract}
Despite remarkable advancements in video depth estimation, existing methods exhibit inherent limitations in achieving geometric fidelity through the affine-invariant predictions, limiting their applicability in reconstruction and other metrically grounded downstream tasks.
We propose {\ourMethod}, a novel framework that recovers high-fidelity point map sequences with temporal coherence from open-world videos, enabling accurate 3D/4D reconstruction, camera parameter estimation, and other depth-based applications. 
At the core of our approach lies a point map Variational Autoencoder (VAE) that learns a latent space agnostic to video latent distributions for effective point map encoding and decoding.
Leveraging the VAE, we train a video diffusion model to model the distribution of point map sequences conditioned on the input videos.
Extensive evaluations on diverse datasets demonstrate that {\ourMethod} achieves state-of-the-art 3D accuracy, temporal consistency, and generalization capability.
%

\end{abstract}

%% file: sections/introduction.tex
\section{Introduction}



Inferring 3D geometry from 2D observations remains a long-standing challenge in computer vision, serving as a fundamental pillar for numerous applications, ranging from autonomous navigation~\cite{dong2022towards,park2021pseudo,wang2019pseudo} and virtual reality~\cite{hong2022depth,zhao2024stereocrafter} to 3D/4D reconstruction~\cite{hu2023tri,xu2024texture,lei2024mosca,wang2024shape} and generation~\cite{yu2024viewcrafter,sun2024dimensionx}. 
However, its inherently ill-posed nature poses persistent difficulties in achieving reliable and consistent geometry estimation from diverse open-world videos.

\begin{figure}[t]
    \centering
    \includegraphics[width=0.99\linewidth]{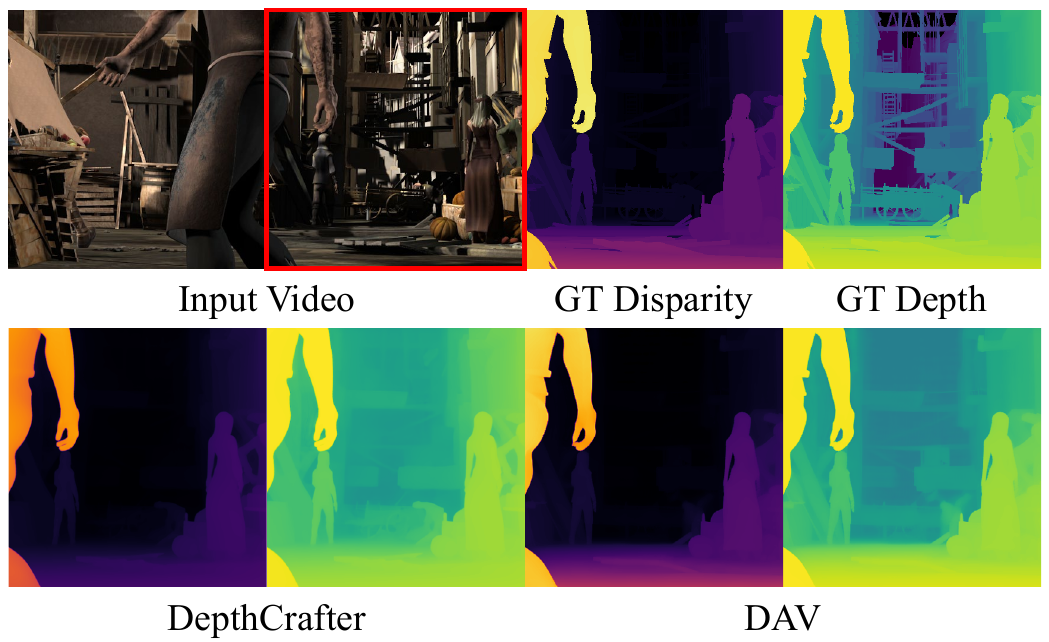}
    \vspace{-0.5em}
     \caption{
        Diffusion-based depth estimation methods, \eg, DepthCrafter~\cite{hu2024-DepthCrafter} and DAV~\cite{yang2024dav}, suffer from significant metric errors in distant regions due to the compression of unbounded depth values into the bounded input range of VAEs.
        }
     \label{fig:problem0}
     \vspace{-1em}
  \end{figure}

Pioneered by Marigold~\cite{marigold}, recent methods harness diffusion models~\cite{ho2020denoising,sohl2015deep,rombach2022sd,blattmann2023svd} to generate affine-invariant depth maps~\cite{he2024lotus,garcia2024e2eft,fu2024geowizard,gui2024depthfm} or sequences~\cite{hu2024-DepthCrafter,yang2024dav,shao2024chrono}, which is achieved by recasting depth cues as pseudo-RGB frames that are suitable for Variational Autoencoder (VAE)~\cite{kingma2013vae} processing.
Although these methods exhibit remarkable spatial and temporal fidelity, the compression of unbounded depth values into the fixed input range of the VAE inevitably leads to a non-trivial information loss, especially for distant scene elements, as shown in~\cref{fig:problem0}.
Moreover, the absence of camera intrinsics and the presence of unknown shift values impede accurate 3D reconstruction, thereby limiting their utility in downstream applications.
Another line of research~\cite{piccinelli2024unidepth,yin2023metric3d,wang2024moge,bochkovskii2024depth,wang2024dust3r} uses pretrained image foundation models to directly estimate metric depth or point maps.
However, neglecting temporal context often induces flickering artifacts when applying these methods to videos.

In this paper, we propose a novel approach, named {\ourMethod}, to estimate high-fidelity and temporally coherent point maps from open-world videos.
These point maps facilitate 3D/4D point cloud reconstruction, camera pose estimation, and the derivation of temporally consistent depth maps and camera intrinsics.
Our method exhibits robust zero-shot generalization capabilities by exploiting the inherent video diffusion priors of natural videos.
Central to our approach is a novel point map VAE, tailored to effectively encode and decode unbounded 3D coordinates without compressing depth values into a bounded range.
It contains a dual-encoder architecture: an encoder inherited from the original video VAE to capture the primary point map information, 
and a newly designed {residual encoder} to embed the remaining information in a latent offset.
Leveraging this design, we can preserve the latent space analogous to the video VAE by regulating the adjusted latent code with the original video decoder.
This analogous latent distribution enables the utilization of pre-trained diffusion weights for robust zero-shot generalizations.
%
%

%

For VAE training, we disentangle the point map into log-space depth and diagonal field of view, rather than directly encoding 3D coordinates in the camera coordinate system or adopting a cuboid-based representation as in prior works~\cite{wang2024dust3r,wang2024moge}.
This disentangled representation demonstrates enhanced suitability for the VAE to capture the intrinsic structure of the point map, largely attributed to its location invariance and resolution independence.
For supervision, we augment the standard reconstruction objective with a normal loss, a multi-scale depth loss to enhance local geometric fidelity, and a regularization term penalizing deviations from the original latent distribution.
Furthermore, our {\ourMethod} integrates a diffusion U-net that generates point map latents from video latents, forming a robust framework for producing high-fidelity and temporally coherent point maps from open-world videos.

We comprehensively evaluate {\ourMethod} on diverse datasets, ranging from static to dynamic scenes, indoor to outdoor environments, and realistic to cartoonish styles.
Our method significantly outperforms existing methods by a large margin, both qualitatively and quantitatively.
Extensive ablation studies validate the effectiveness of our proposed components, and demonstrate the applicability of our method to 3D/4D point cloud reconstruction and camera pose estimation.
Our contributions are summarized as follows:
\begin{itemize}
    \item We present {\ourMethod}, a novel approach for estimating high-fidelity and temporally coherent geometry from diverse open-world videos.
    \item We propose a point map VAE for effective encoding and decoding of point maps, which employs a dual-encoder architecture to maintain the latent space analogous to the inherited video VAE for generalization ability.
    \item We introduce the disentangled point map representation and multi-scale depth loss to train the VAE, significantly improving the robustness and fidelity of our method.
\end{itemize}


%% file: sections/related_works.tex
\section{Related Works}

\begin{figure*}[t]
    \centering
    \includegraphics[width=0.99\linewidth]{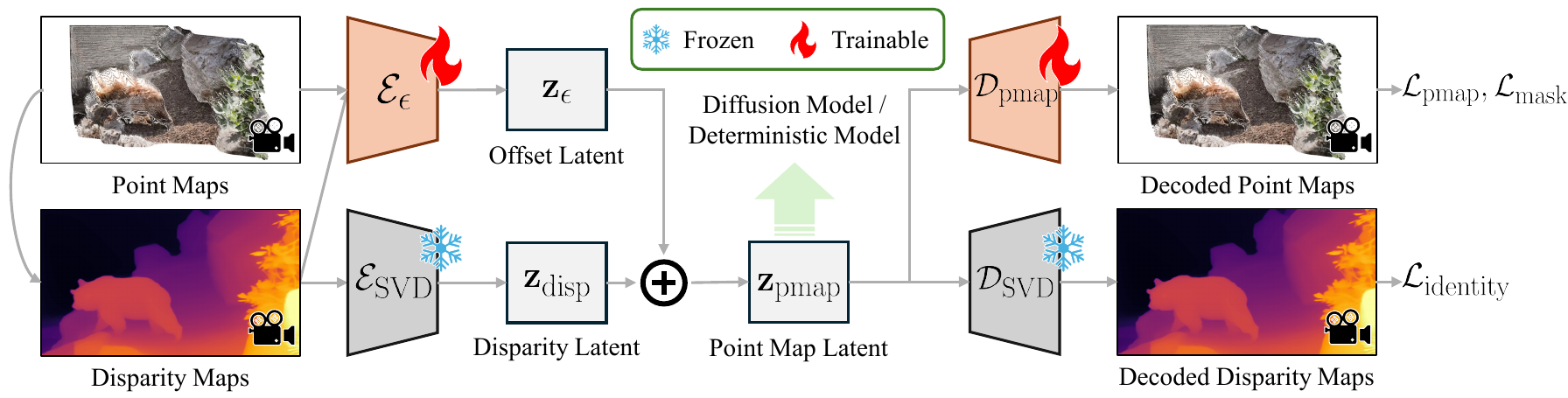}
    \vspace{-0.5em}
     \caption{\textbf{Architecture of our point map VAE.} 
      The point map VAE encodes and decodes point maps with unbounded values, alleviating the inaccurate prediction in distant regions. 
      We adopt a dual-encoder design: the native encoder $\mathcal{E}_\text{SVD}$ inherited from SVD captures normalized disparity maps, while a residual encoder $\mathcal{E}_\epsilon$ embeds remaining information as an offset.
      It preserves the original latent space by regulating the latents via the original decoder $\mathcal{D}_\text{SVD}$, enabling the utilization of pretrained diffusion priors.
      A point map decoder $\mathcal{D}_\text{pmap}$ recovers the final point maps from the latent codes.
     }
     \vspace{-1em}
     \label{fig:vae}
  \end{figure*}

\boldstartspace{Monocular depth estimation (MDE).} 
MDE methods~\cite{bhat2021adabins,eigen2014depth,li2024binsformer,fu2018deep,lee2019big,aich2021bidirectional,li2023depthformer,yang2021transformer,patil2022p3depth} predict depth maps from single images or videos.
To achieve zero-shot generalization, MiDaS~\cite{ranftl2020midas} introduces affine-invariant supervision and trains on mixing datasets.
Depth Anything~\cite{depthanything} and its V2~\cite{depth_anything_v2} extend this framework to transformer-based architectures~\cite{oquab2023dinov2} and semi-supervised learning, using large-scale unlabeled images for improved generalization.
Pioneered by Marigold~\cite{marigold}, recent works~\cite{gui2024depthfm, fu2024geowizard, pham2024sharpdepth, he2024lotus, garcia2024e2eft} adapt pretrained image diffusion models~\cite{rombach2022sd} to MDE by converting depth maps to pseudo-RGB representations and exclusively finetuning the U-net on depth latent codes, achieving superior quality and robustness.
To generalize to videos, previous methods~\cite{chen2019self, kopf2021rcvd, Luo-VideoDepth-2020, zhang2021consistent, yasarla2023mamo, wang2023neural} employ test-time optimization, memory mechanisms, or stabilization networks for temporal coherence, whereas recent studies~\cite{hu2024-DepthCrafter, yang2024dav, shao2024chrono} finetune video diffusion models~\cite{blattmann2023svd} to yield high-quality, temporally consistent depth sequences.
However, these methods ignore the camera intrinsic estimation and provide only affine-invariant depth, which is scale and shift ambiguous, hindering 3D accuracy and downstream applications that require projection into 3D space.

\boldstartspace{Monocular geometry estimation (MGE).}
To overcome these limitations, MGE methods jointly infer camera parameters and metric or up-to-scale depth maps.
LeRes~\cite{yin2021learning, yin2022towards} utilizes 3D point cloud encoders to recover missing shift and focal parameters during depth estimation.
UniDepth~\cite{piccinelli2024unidepth} decouples camera parameter prediction from depth estimation via a pseudo-spherical 3D representation and a camera self-prompting mechanism.
DepthPro~\cite{bochkovskii2024depth} introduces a ViT-based design~\cite{oquab2023dinov2} for high-resolution depth estimation, coupled with a dedicated image encoder for focal length prediction.
DUSt3R~\cite{wang2024dust3r} projects two-view images or identical pairs to scale-invariant point maps in camera space, facilitating the derivation of camera intrinsic and depth maps.
MoGe~\cite{wang2024moge} employs affine-invariant point maps to mitigate focal-distance ambiguity, achieving state-of-the-art performance.
%
Besides, Metric3D~\cite{yin2023metric3d} and its V2~\cite{hu2024metric3d} rely on user-provided camera parameters to estimate metrically accurate depth maps.
However, these approaches are restricted to static images and incur flickering artifacts when directly applied to video sequences.

\boldstartspace{Steganography and information hiding.} 
%
Steganography visually hides secret information within existing features~\cite{baluja2017hiding, jing2021hinet, zhu2018hidden}.
Previous works have demonstrated the capacity of neural networks to embed visual data within images or videos, such as invertible down-scaling~\cite{xiao2020invertible}, gray-scaling~\cite{xia2018invertible}, and mono-nizing~\cite{hu2020mononizing}.
The most relevant work to ours is LayerDiffuse~\cite{zhang2024layerdiffuse}, which conceals image transparency information within a small perturbation in the latent space of Stable Diffusion~\cite{rombach2022sd}.
Adhering to the same insight, we encode point maps into the latent space of diffusion models while preserving the underlying distribution, facilitating the utilization of pretrained diffusion models for geometry estimation.

%% file: sections/method.tex
\section{Method}

Given an input RGB video $\mathbf{v} \in \mathbb{R}^{T\times H \times W \times 3}$, we aim to predict a temporally consistent point map sequence $\mathbf{p} \in \mathbb{R}^{T\times H \times W \times 3}$ alongside a valid mask $\mathbf{m} \in [0,1]^{T\times H \times W}$ to exclude undefined regions (\eg, sky).
Each point map contains the 3D coordinates $p=(x_\mathbf{p}, \, y_\mathbf{p}, \, z_\mathbf{p})^T$ in the camera coordinate system for every pixel.
To this end, we propose {\ourMethod}, a novel approach that leverages video diffusion models (VDMs) for robust point map estimation from open-world videos.
We model the joint distribution $\mathcal P(\mathbf{p}, \, \mathbf{m} \, | \, \mathbf{v})$ in the latent space.
While VDMs’ native VAE effectively encodes video frames and masks, accurate point map representation necessitates a dedicated VAE tailored for geometric encoding and decoding.

\subsection{Architecture of Point Map VAE}
\label{sec:vae}

Existing diffusion-based depth estimation methods~\cite{hu2024-DepthCrafter,yang2024dav} simply employ the native VAE to encode and decode only partial information from the point maps, \ie the normalized disparity maps $\widetilde{\mathbf{x}}_\text{disp}$:
\begin{equation}
\begin{aligned}
    \widetilde{\mathbf{x}}_\text{disp} &= 2 \times \frac{\mathbf{x}_\text{disp} - \min(\mathbf{x}_\text{disp})}{\max(\mathbf{x}_\text{disp}) - \min(\mathbf{x}_\text{disp})} - 1, \\
    \mathbf{x}_\text{disp} &= {b\cdot f} \, / \, {z_\mathbf{p}},
\end{aligned}
\label{eq:normalization}
\end{equation}
where $b$ is the baseline, $f$ is the focal length, and $z_\mathbf{p}$ is the z-coordinate of the point map $\mathbf{p}$.
However, such normalization often misestimates depths in distant regions (\cref{fig:problem0}), resulting in geometric distortions due to compressing unbounded depths into the VAE's fixed input range.

To this end, we propose a point map VAE that directly handles point maps over the unbounded range $[0, +\infty]$.
Crucially, its latent distribution should be tightly aligned with that of the native VAE to fully exploit pre-trained VDMs.
Inspired by LayerDiffuse~\cite{zhang2024layerdiffuse}, we propose a dual-encoder architecture: the inherited native encoder $\mathcal{E}_\text{SVD}$ captures the primary point map features, while a newly designed residual encoder $\mathcal{E}_\epsilon$ encodes remaining information as an offset (see~\cref{fig:vae}).  
Given that the normalized disparity maps $\widetilde{\mathbf{x}}_\text{disp}$ encapsulate significant relative depth cues, we employ $\mathcal{E}_\text{SVD}$ on $\widetilde{\mathbf{x}}_\text{disp}$ and harness $\mathcal{E}_\epsilon$ to embed the residual information into the offset.
The final point map latent is obtained by their summation:
\begin{equation}
\begin{aligned}
    \mathbf{z}_\text{pmap} = \mathcal{E}_\text{SVD}(\widetilde{\mathbf{x}}_\text{disp}) + \mathcal{E}_\epsilon(\mathbf{p}, \mathbf{m}, \widetilde{\mathbf{x}}_\text{disp}),
\end{aligned}
\end{equation}
%
This dual-encoder architecture allows us to explicitly regularize the latent space of $\mathbf{z}_\text{pmap}$ to avoid disrupting the original latent distribution.
Considering most VAEs in VDM are diagonal Gaussian models (\ie mean and variance), we apply the offset solely to the mean, retaining the original variance for simplicity.
For decoding, we design a dedicated decoder $\mathcal{D}_\text{pmap}$ to reconstruct both the point map $\widehat{\textbf{p}}$ and the valid mask $\widehat{\textbf{m}}$:
\begin{equation}
    \widehat{\textbf{p}},\widehat{\textbf{m}} = \mathcal D_\text{pmap}(\mathbf{z}_\text{pmap}).
\end{equation}
To ensure temporal consistency, we employ temmporal layers in the decoder to capture the temporal dependencies across frames.

\subsection{Training of Point Map VAE}
\label{sec:repr}
\boldstartspace{Point map representation.}
The points $\mathbf{p} = (x_\mathbf{p}, \, y_\mathbf{p}, \, z_\mathbf{p})^T$ are scattered non-uniformly across the view frustum, resulting in a complex spatial distribution that poses a challenge to deep networks in capturing their inherent structure.
To mitigate this, existing point map estimation methods~\cite{wang2024dust3r,wang2024moge} assume a centered camera principal point and remap depth values into log-space, thereby projecting the points into a cuboid domain:
\begin{equation}
\begin{aligned}
    \mathbf{p}_\text{cuboid} = \left[{x_\mathbf{p}}/{z_\mathbf{p}}, \; {y_\mathbf{p}}/{z_\mathbf{p}}, \; \log z_\mathbf{p}\right].
\end{aligned}
\label{eq:raw_pmap}
\end{equation}
However, this representation is suboptimal for our point map VAE.
In particular, the first two channels of $\mathbf{p}_\text{cuboid}$ encode ray directions from the camera center to each pixel, conveying location-specific information that diverges from the translation-invariant nature of RGB features.
To address this discrepancy, we propose decoupling the point map into:
\begin{equation}
\begin{aligned}
    \mathbf{p}_\text{dec} &= \left[\theta_\text{diag}, \; \log z_\mathbf{p}\right], 
\end{aligned}
\label{eq:pmap}
\end{equation}
where $\theta_\text{diag}={\sqrt{W^2 + H^2}}\;/\;{2f}$ denotes the diagonal field of view, a constant map for all points in a frame.
Since $\mathbf{p}_\text{dec}$ is independent of spatial location, it is more suitable for our VAE to learn an effective latent distribution.
Moreover, this formulation enables us to train our network only on fixed-resolution videos, while generalizing to varying resolutions and aspect ratios, owing to the invariance of $\theta_\text{diag}$.
The original point map $\mathbf{p}$ can be effortlessly recovered from $\mathbf{p}_\text{dec}$ via the inverse perspective transformation.

\begin{figure}[!t]
    \centering
    \includegraphics[width=0.99\linewidth]{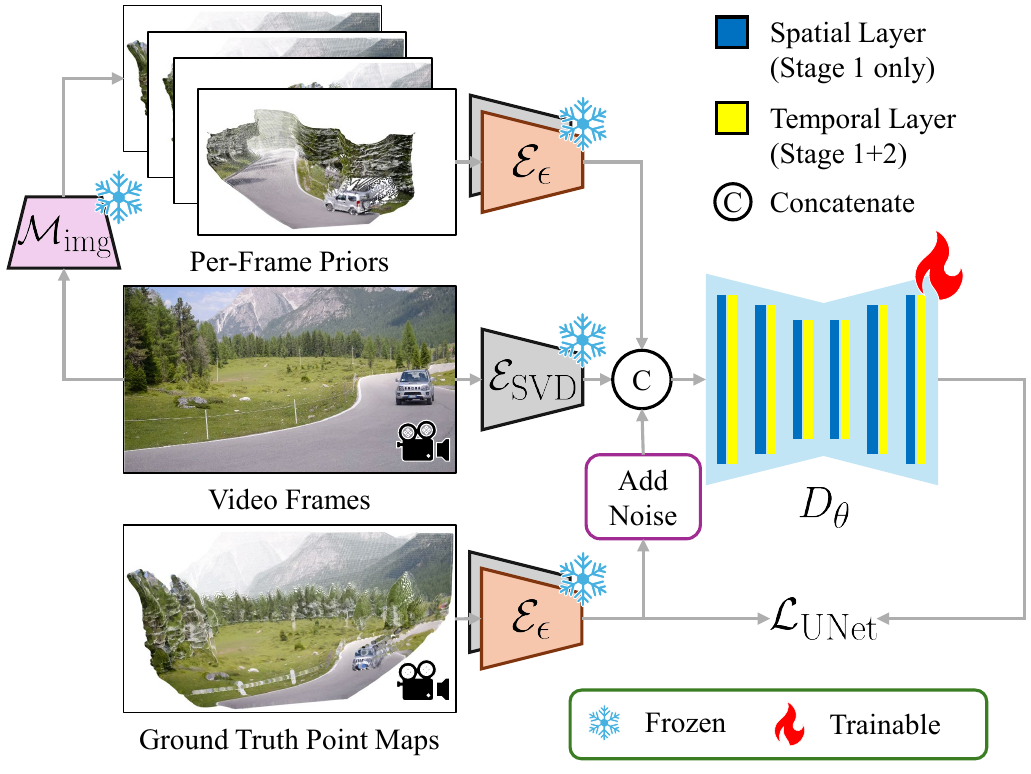}
      \vspace{-0.5em}
     \caption{
        \textbf{Diffusion UNet.}
        We jointly condition the diffusion model on video latents and per-frame geometry priors from an image MGE model $\mathcal{M}_\text{img}$.
        The geometry is encoded into latent space via our point map VAE, while the video latents are obtained from the native VAE.
        }
      \vspace{-1em}
     \label{fig:unet}
  \end{figure}

\boldstartspace{Loss functions.}
To train the point map VAE, we define reconstruction loss $\mathcal{L}_\text{recon}$ as the $L_1$ norm between the decoded depth and diagonal field of view and their ground truth counterparts.
Besides, we also impose a mask loss $\mathcal{L}_\text{mask}$ to exclude undefined regions, \eg sky, as the $L_2$ norm between the predicted and ground truth valid masks.
To promote surface quality, we introduce a normal loss $\mathcal{L}_\text{n}$ that supervises the normal maps derived from the reconstructed point maps and the ground truth, as well as a multi-scale depth loss $\mathcal{L}_\text{ms}$ that measures the alignment between reconstructed and ground truth depth maps within local regions, inspired by MoGe~\cite{wang2024moge}.
Importantly, to regularize our latent space agnostic to the original SVD's latent distribution, we employ a loss term $\mathcal{L}_\text{identity}$ to penalize the latent deviation:
\begin{equation}
    \begin{aligned}
        \mathcal{L}_\text{identity} &= ||\widetilde{\mathbf{x}}_\text{disp} - \mathcal D_\text{SVD}(\mathbf{z}_\text{pmap})||_2^2.
    \end{aligned}
\end{equation}
The final training objective $\mathcal{L}_\text{VAE}$ is defined as:
\begin{equation}
    \begin{aligned}
        \mathcal{L}_\text{VAE} &=   \underbrace{\mathcal{L}_\text{recon} + \mathcal{L}_\text{ms} + \lambda_\text{n}\mathcal{L}_\text{n}}_{\mathcal L_\text{pmap}} + \mathcal{L}_\text{identity} + \lambda_\text{mask}\mathcal{L}_\text{mask}. 
    \end{aligned}
\end{equation}
Please refer to the supplementary material for more details on the loss functions.

\begin{table*}[t]
\centering
\caption{
    \textbf{Evaluation on point map estimation.} 
    Results are aligned with the ground truth by optimizing a shared scale factor across the entire video.
    $\text{Rel}^p$ and $\delta^p$ are in percentage. The best and second-best results are highlighted in \textbf{bold} and \underline{underline}, respectively. ``G'' denotes the diffusion version of our model and ``D'' denotes the deterministic variant.
    }
\vspace{-0.5em}
\label{tab:comp_point}
\begin{adjustbox}{width=\textwidth}
\begin{tabular}{l|cc|cc|cc|cc|cc|cc|cc|c}
\toprule
\multirow{2}{*}{Method} & \multicolumn{2}{c|}{GMU Kitchen~\cite{georgakis2016multiview}} & \multicolumn{2}{c|}{Monkaa~\cite{mayer2016large}} & \multicolumn{2}{c|}{Sintel~\cite{sintel}} & \multicolumn{2}{c|}{ScanNet~\cite{dai2017scannet}} & \multicolumn{2}{c|}{DDAD~\cite{ddad2020}} & \multicolumn{2}{c|}{KITTI~\cite{Geiger2013IJRR}} & \multicolumn{2}{c|}{DIODE~\cite{diode_dataset}} & Avg. \\

& $\text{Rel}^{p}\!\downarrow$ & $\delta^{p}\!\uparrow$ & $\text{Rel}^{p}\!\downarrow$ & $\delta^{p}\!\uparrow$ & $\text{Rel}^{p}\!\downarrow$ & $\delta^{p}\!\uparrow$ & $\text{Rel}^{p}\!\downarrow$ & $\delta^{p}\!\uparrow$ & $\text{Rel}^{p}\!\downarrow$ & $\delta^{p}\!\uparrow$ & $\text{Rel}^{p}\!\downarrow$ & $\delta^{p}\!\uparrow$ & $\text{Rel}^{p}\!\downarrow$ & $\delta^{p}\!\uparrow$ & Rank $\!\downarrow$ \\
\midrule
DUSt3R~\cite{wang2024dust3r}$^\dagger$ & 22.2 & 68.2 & 37.0 & 45.1 & 43.9 & 35.6 & \notzs{13.3} & \notzs{87.7} & 37.8 & 37.3 & 17.2 & 87.8 & 20.0 & 85.3 & 6.3 \\
MonST3R~\cite{zhang2024monst3r}$^{\dagger}$ & 24.1 & 64.4 & 40.2 & 32.6 & 40.0 & 34.1 & \notzs{13.6} & \notzs{87.2} & 40.7 & 25.9 & 24.0 & 58.1 & 22.2 & 79.4 & 7.4 \\
MonST3R~\cite{zhang2024monst3r}$^{\ddagger}$ & 11.4 & 91.5 & 36.2 & 42.6 & 38.6 & 35.9 & \notzs{6.13} & \notzs{97.6} & 40.0 & 29.3 & 25.0 & 58.7 & 22.2 & 79.4 & 5.4 \\
UniDepth~\cite{piccinelli2024unidepth} & 9.96 & 94.1 & 23.0 & 65.6 & 30.9 & 50.6 & \notzs{6.54} & \notzs{98.4} & 23.0 & 64.9 & \first{4.24} & \first{99.3} & 16.1 & 88.0 & 2.9 \\
DepthPro~\cite{bochkovskii2024depth} & 14.0 & 86.5 & 29.5 & 50.4 & 45.0 & 36.3 & \notzs{10.5} & \notzs{93.6} & 39.8 & 43.1 & 12.8 & 93.6 & 18.6 & 87.1 & 5.6 \\
MoGe~\cite{wang2024moge} & 21.3 & 69.1 & 28.0 & 58.1 & 31.2 & 52.0 & 13.5 & 88.0 & 16.0 & 85.5 & 8.51 & 95.7 & 13.5 & \first{93.5} & 4.4 \\
\midrule
Ours(D) & \second{8.88} & \first{94.3} & \first{18.8} & \first{79.7} & \second{25.9} & \second{62.8} & \first{8.92} & \first{96.4} & \second{15.6} & \second{89.0} & 6.73 & 98.4 & \first{13.0} & \second{92.8} & \second{2.0} \\
Ours(G) & \first{8.52} & \first{94.3} & \second{20.5} & \second{75.5} & \first{25.6} & \first{64.9} & \second{9.16} & \second{96.0} & \first{15.0} & \first{90.6} & \second{6.34} & \second{98.7} & \second{13.1} & 92.7 & \first{1.9} \\
\bottomrule
\toprule
& $\text{Rel}^{d}\!\downarrow$ & $\delta^{d}\!\uparrow$ & $\text{Rel}^{d}\!\downarrow$ & $\delta^{d}\!\uparrow$ & $\text{Rel}^{d}\!\downarrow$ & $\delta^{d}\!\uparrow$ & $\text{Rel}^{d}\!\downarrow$ & $\delta^{d}\!\uparrow$ & $\text{Rel}^{d}\!\downarrow$ & $\delta^{d}\!\uparrow$ & $\text{Rel}^{d}\!\downarrow$ & $\delta^{d}\!\uparrow$ & $\text{Rel}^{d}\!\downarrow$ & $\delta^{d}\!\uparrow$ & Rank $\!\downarrow$ \\
\midrule
DUSt3R~\cite{wang2024dust3r}$^{\dagger}$ & 21.6 & 64.0 & 35.8 & 41.9 & 41.3 & 36.0 & \notzs{13.1} & \notzs{84.5} & 32.3 & 46.5 & 10.9 & 87.9 & 15.9 & 85.4 & 6.7 \\
MonST3R~\cite{zhang2024monst3r}$^{\dagger}$ & 22.6 & 61.9 & 38.8 & 31.4 & 37.5 & 33.4 & \notzs{13.3} & \notzs{84.4} & 30.7 & 46.4 & 9.06 & 91.8 & 17.4 & 80.8 & 6.6 \\
MonST3R~\cite{zhang2024monst3r}$^{\ddagger}$ & 8.91 & 90.7 & 33.9 & 41.3 & 35.9 & 36.1 & \notzs{5.18} & \notzs{97.1} & 31.9 & 46.1 & 13.4 & 80.4 & 17.4 & 80.8 & 5.3 \\
UniDepth~\cite{piccinelli2024unidepth}& \first{8.11} & \first{93.7} & 20.8 & 60.0 & 28.4 & 48.5 & \notzs{5.32} & \notzs{98.0} & 22.9 & 63.4 & \first{3.45} & \first{99.1} & 11.5 & 89.9 & 2.6 \\
DepthPro~\cite{bochkovskii2024depth} & 14.0 & 83.1 & 28.4 & 45.2 & 43.2 & 34.3 & \notzs{10.0} & \notzs{90.7} & 38.3 & 40.6 & 9.47 & 91.5 & 12.6 & 87.4 & 6 \\
MoGe~\cite{wang2024moge}& 20.6 & 64.7 & 25.7 & 54.8 & 29.2 & 49.0 & 13.3 & 84.9 & 14.6 & 85.2 & 7.69 & 94.1 & \first{8.13} & \first{93.5} & 4.3 \\
\midrule
Ours(D) & 8.51 & 93.4 & \first{16.1} & \first{77.1} & \first{22.2} & \first{65.7} & \first{7.88} & \first{95.5} & \second{12.3} & \second{88.4} & 5.88 & 97.6 & 10.2 & 92.1 & \second{2.3} \\
Ours(G) & \second{8.30} & \second{93.6} & \second{18.3} & \second{71.5} & \second{22.6} & \second{63.7} & \second{8.39} & \second{95.0} & \first{12.0} & \first{90.4} & \second{5.44} & \second{98.2} & \second{10.0} & \second{92.4} & \first{2.1} \\
\bottomrule
\end{tabular}
\end{adjustbox}
\vspace{1em}
{\footnotesize{
    $^*$: Not strictly zero-shot (trained on ScanNet~\cite{dai2017scannet} or ScanNet++~\cite{yeshwanth2023scannet++});
    $^\dagger$: Inference with duplicated frames;
    $^\ddagger$: Post-optimization with external data.
}}
\vspace{-2em}
\end{table*}

\subsection{Diffusion UNet}
\label{sec:unet}
%

Since our point map VAE is meticulously designed and regularized to align closely with the original SVD's latent distribution, we can train a diffusion UNet to estimate point maps from videos with only synthetic data, using the pretrained generative prior of video diffusion models.
Although it significantly alleviates the issue of lacking high-quality point map annotations in real-world videos, the synthetic data still suffers from limited diversity in camera intrinsics, which may degrade the generalization ability of diagonal field-of-view prediction on real-world scenarios.
%
%
To mitigate this, alongside video latents, we propose the integration of per-frame geometry priors as conditioning inputs within the diffusion UNet, as shown in ~\cref{fig:unet}.
%
We employ our point map VAE to encode the per-frame point maps predicted by MoGe~\cite{wang2024moge} into the latent space to act as geometry priors that provide strong camera-intrinsic clues, although they may suffer from inaccuracies and flickering.

Following DepthCrafter~\cite{hu2024-DepthCrafter}, we train the diffusion UNet with the EDM~\cite{karras2022edm} pre-conditioning and noise schedule, and adopt a multi-stage training strategy to capture long temporal context under GPU memory constraints.
After training, the UNet can process videos with varying lengths (\eg 1 to 110 frames) at a time, and we adopt the stitching inference strategy~\cite{hu2024-DepthCrafter} to handle videos with arbitrary lengths.
Besides, inspired by recent advancements~\cite{garcia2024e2eft,he2024lotus} in reformulating the diffusion process into a deterministic single-step framework for depth estimation, we also train a deterministic variant by removing the noisy latent from input.

%% file: sections/experiments.tex
\begin{figure*}[!t]
  \centering
  \includegraphics[width=0.99\linewidth]{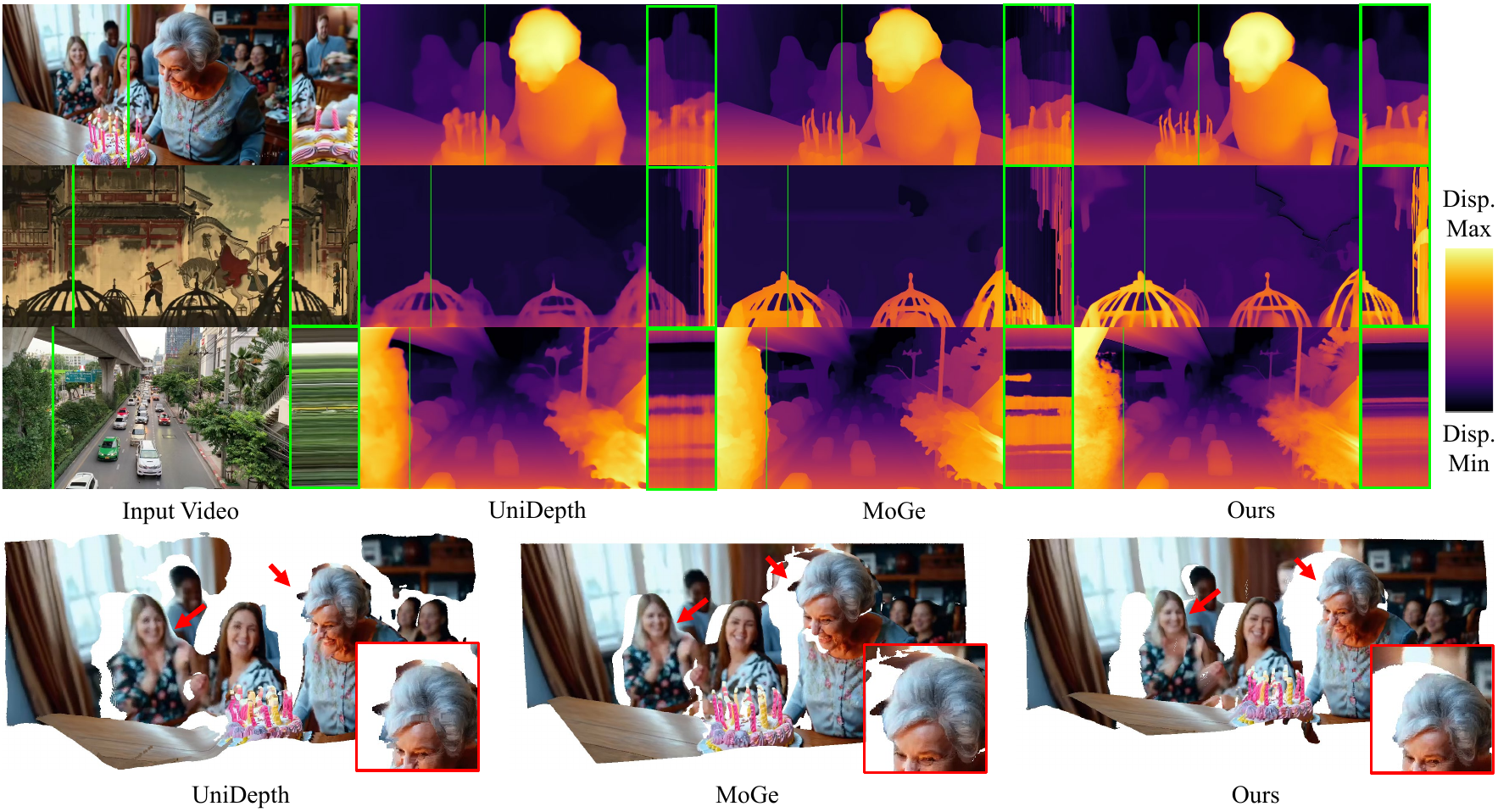}
  \vspace{-1em}
   \caption{
    \textbf{Qualitative comparison of point map estimation.} 
    Disparity maps are derived from estimated point maps via~\cref{eq:normalization}.
    The green boxes highlight temporal profiles of the disparity maps, sliced along the time axis at the green lines.
    Zoom in for better visualization.
    }
    \vspace{-0.5em}
   \label{fig:comp_point}
\end{figure*}

\begin{table*}[t]
\centering
  \caption{
    \textbf{Evaluation on depth map estimation.} 
      Results are aligned with the ground truth by optimizing a shared scale factor and shift across the entire video.
    $\text{Rel}^d$ and $\delta^d$ are in percentage. The best and second-best results are highlighted in \textbf{bold} and \underline{underline}, respectively. ``G'' denotes the diffusion version of our model and ``D'' denotes the deterministic variant.
    }
\vspace{-0.5em}
  \label{tab:comp_depth}
\begin{adjustbox}{width=\textwidth}
  \begin{tabular}{l|cc|cc|cc|cc|cc|cc|cc|c}
  \toprule
  \multirow{2}{*}{Method} & \multicolumn{2}{c|}{GMU Kitchen~\cite{georgakis2016multiview}} & \multicolumn{2}{c|}{Monkaa~\cite{mayer2016large}} & \multicolumn{2}{c|}{Sintel~\cite{sintel}} & \multicolumn{2}{c|}{Scannet~\cite{dai2017scannet}} & \multicolumn{2}{c|}{DDAD~\cite{ddad2020}} & \multicolumn{2}{c|}{KITTI~\cite{Geiger2013IJRR}} & \multicolumn{2}{c|}{DIODE~\cite{diode_dataset}} & Avg. \\
  & $\text{Rel}^{d}\!\downarrow$& $\delta^{d}\!\uparrow$ & $\text{Rel}^{d}\!\downarrow$ & $\delta^{d}\!\uparrow$ & $\text{Rel}^{d}\!\downarrow$ & $\delta^{d}\!\uparrow$ & $\text{Rel}^{d}\!\downarrow$ & $\delta^{d}\!\uparrow$ & $\text{Rel}^{d}\!\downarrow$ & $\delta^{d}\!\uparrow$ & $\text{Rel}^{d}\!\downarrow$ & $\delta^{d}\!\uparrow$ & $\text{Rel}^{d}\!\downarrow$ & $\delta^{d}\!\uparrow$ & Rank\small{↓} \\
  \midrule
  DA~\cite{depthanything} & 18.0 & 68.2 & 24.1 & 62.1 & 37.2 & 59.8 & 11.3 & 87.8 & 13.4 & 85.1 & 8.70 & 92.4 & \first{6.94} & 94.6 & 4.1 \\
  DA V2~\cite{depth_anything_v2} & 19.3 & 65.4 & 24.0 & 61.5 & 40.6 & 55.0 & 12.3 & 85.1 & 13.9 & 84.7 & 11.1 & 87.0 & \first{6.94} & 95.0 & 5.1 \\
  ChronoDepth~\cite{shao2024chrono} & 20.1 & 65.2 & 35.1 & 52.6 & 45.1 & 57.9 & 14.3 & 81.4 & 34.6 & 45.8 & 15.0 & 79.8 & 12.2 & 90.6 & 6.9   \\
  DepthCrafter~\cite{hu2024-DepthCrafter} & 13.8 & 80.6 & 23.4 & 73.8 & 30.5 & 67.0 & 11.3 & 87.3 & 15.6 & 80.7 &  9.96 & 89.6 & 12.6 & 86.2 & 4.9  \\
  DAV~\cite{yang2024dav} & 10.8 & 89.4 & 19.0 & 72.3 & 35.7 & 67.3 & 8.83 & 92.7 & \first{12.3} & 85.4 & 7.13 & 95.3 & 7.47 & 93.7 & 3.1  \\
  \midrule
  Ours(D) & \second{8.28} & \second{93.2} & \first{12.0} & \first{83.5} & \first{16.3} & \first{74.3} & \first{7.27} & \first{96.1} & 13.4 & \second{86.2} & \second{5.60} & \second{97.7} & 7.00 & \first{96.2} &  \first{1.9}\\
  Ours(G) & \first{8.03} & \first{94.0} & \second{13.0} & \second{80.5} & \second{16.9} & \second{73.2} & \second{7.57} & \second{95.9} & \second{12.7} & \first{87.5} & \first{5.25} & \first{98.3} & 7.03 & \second{96.1} &  \second{2.0} \\
  \bottomrule
  \end{tabular}
\end{adjustbox}
  \vspace{-1em}
\end{table*}

\section{Experiments}

\subsection{Implementation Details}


We build {\ourMethod} upon the SVD~\cite{blattmann2023svd} framework.
The residual encoder and point decoder in the point map VAE adopt the same architecture as in SVD's VAE, supplemented by zero convolution~\cite{zhang2023adding} in the output layers.
We collected 14 synthetic RGBD datasets~\cite{wang2020gtasfm,roberts2021hypersim,wang2021irs,niklaus20193d,li2023matrixcity,fonder2019mid,huang2018deepmvs,mehl2023spring,zheng2020structured3d,ros2016synthia,wang2020tartanair,gomez2023all,cabon2020virtual,karaev2023dynamicstereo}, comprising \textbf{1.85M} frames, for training.
Among these, 11 datasets can form \textbf{12K} video clips with up to 150 frames each.
For training stability, we normalize point clouds with a shared scale factor across frames, yielding up-to-scale point clouds akin to structure-from-motion~\cite{schonberger2016structure}.
We first train the point map VAE from scratch on RGBD images with an AdamW~\cite{loshchilov2017decoupled} optimizer at a learning rate of $10^{-4}$ for 40K iterations, then finetune on video data for an additional 20K iterations. 
The diffusion UNet is finetuned with a learning rate of $10^{-5}$ for 40K and 30K iterations in two stages.
All experiments are conducted on 8 GPUs and take about 3 days.
Further details are in the supplementary material.

\subsection{Quantitative and Qualitative Evaluation}

\noindent\textbf{Evaluation protocol.} 
For evaluation, we employ seven datasets unseen during training:
\textbf{GMU Kitchens}~\cite{georgakis2016multiview} and \textbf{ScanNet}~\cite{dai2017scannet} are captured with Kinect for indoor scenes;
\textbf{DDAD}~\cite{ddad2020} and \textbf{KITTI}~\cite{Geiger2013IJRR} are collected via lidar sensors for outdoor driving;
\textbf{Monkaa}~\cite{mayer2016large} and \textbf{Sintel}~\cite{sintel} are synthetic datasets with precise depth annotations and challenging dynamics;
and \textbf{DIODE}~\cite{diode_dataset} is a high-resolution image dataset with far-range depth maps.
Besides, we also qualitatively evaluate on DAVIS~\cite{Perazzi2016davis}, DL3DV~\cite{ling2024dl3dv}, Sora~\cite{sora}-generated, and open-world videos.
To assess up-to-scale point map quality, we use the relative point error $\text{Rel}^p$ and percentage of inliers $\delta^{p}$ (threshold $0.25$), following MoGe~\cite{wang2024moge}.
We align predicted point maps with ground truth by optimizing \emph{a shared scale factor across the entire video} for all methods.
%
%
We also evaluate derived depth sequences using the absolute relative error $\text{Rel}^d$ and the inlier percentage $\delta^d$ (threshold $1.25$), following~\cite{hu2024-DepthCrafter}.

\begin{figure*}[!t]
  \centering
  \includegraphics[width=0.99\linewidth]{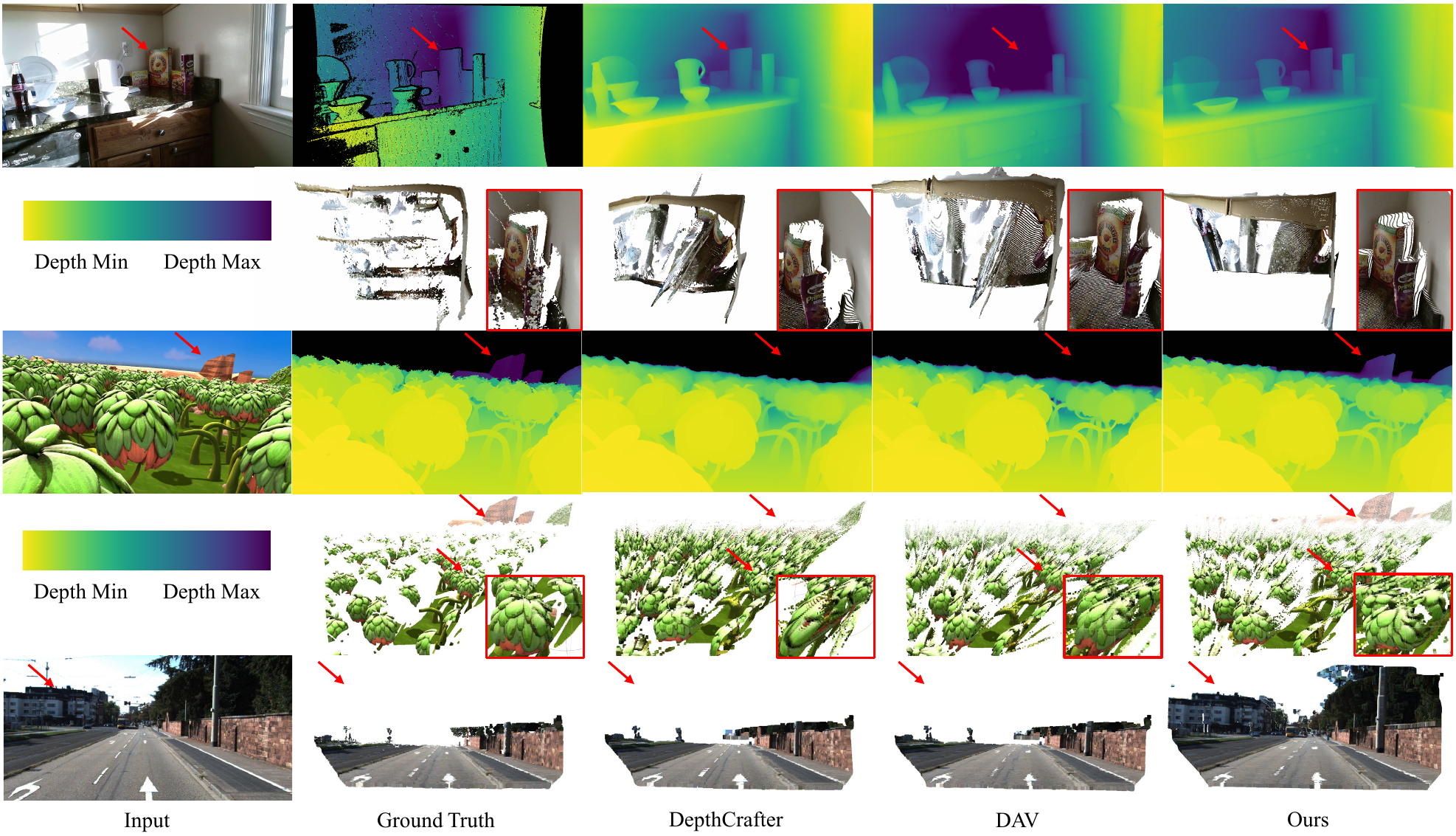}
    \vspace{-0.7em}
   \caption{\textbf{Qualitative comparison of depth map estimation.} We transform point maps and disparity maps into metric depth maps for better visualization of distant regions. Zoom in for better visualization.}
   \vspace{-1.2em}
   \label{fig:comp_depth}  
\end{figure*}

\noindent\textbf{Evaluation on point maps.}
We compare our method with representative point map estimation approaches, \eg, DUSt3R~\cite{wang2024dust3r}, MonST3R~\cite{zhang2024monst3r}, UniDepth~\cite{piccinelli2024unidepth}, DepthPro~\cite{bochkovskii2024depth}, and MoGe~\cite{wang2024moge}.
Among them, DUSt3R and MonST3R are designed for two-view scenarios, addressing static and dynamic scenes, respectively, and are evaluated by inputting two identical frames.
For MonST3R, we also evaluate with its post-processing, which requires external optical flows to refine global point clouds and poses.
As shown in~\cref{tab:comp_point}, our method outperforms others on most benchmarks, with substantial gains on the challenging Monkaa and Sintel datasets.
Although UniDepth shows better performance on KITTI (likely due to training on DrivingStereo~\cite{yang2019drivingstereo} with a shared LiDAR sensor), our approach attains a superior average rank.
For the image benchmark DIODE, our method still achieves competitive performance compared to methods specialized for static images.
Notably, some methods are trained on ScanNet~\cite{dai2017scannet} or ScanNet++~\cite{yeshwanth2023scannet++}, violating the zero-shot evaluation, yet our method sustains comparable accuracy on ScanNet.
Moreover, visual comparisons in~\cref{fig:comp_point} indicate that only our method can produce temporally consistent point maps with fine-grained details, while others (UniDepth and MoGe) exhibit issues like flickering or blurred details.

\begin{figure}[t]
  \centering
  \includegraphics[width=0.99\linewidth]{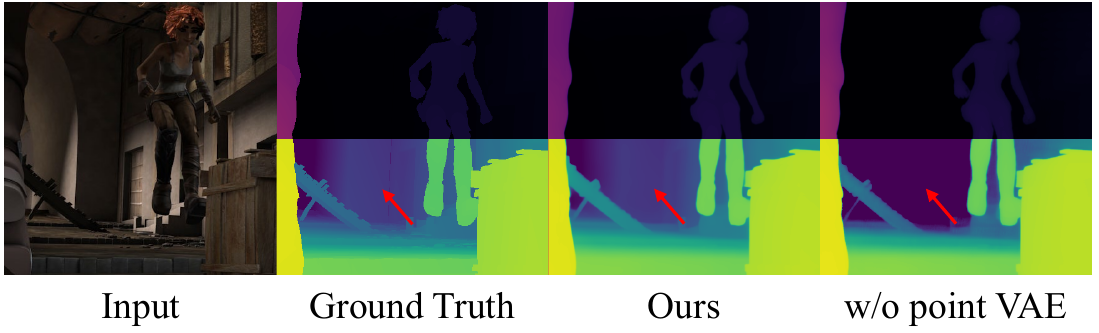}
  \vspace{-0.8em}
   \caption{
    Comparison on disparity (top) and depth (bottom) quality between our full model and the w/o point map VAE variant.
    }
    \vspace{-1.2em}
   \label{fig:ablate_depth}
\end{figure}


\noindent\textbf{Evaluation on depth maps.}
To compare our method with cutting-edge monocular depth estimation methods, \eg, ChronoDepth~\cite{shao2024chrono}, DepthCrafter~\cite{hu2024-DepthCrafter}, DAV~\cite{yang2024dav}, and DepthAnything (DA) V1~\cite{depthanything} and V2~\cite{depth_anything_v2}, we follow the evaluation protocol in~\cite{hu2024-DepthCrafter}, except for a center crop to meet the aspect ratio requirement (0.5 to 2).
As shown in~\cref{tab:comp_depth}, our method achieves the best performance on almost all video datasets and remains competitive even on the image dataset DIODE.
The qualitative comparison in~\cref{fig:comp_depth} demonstrates that our method generates superior depth maps and point clouds, \eg, the potato chip bucket and the plant in the first two examples.
In driving scenarios, such as the third example, DepthCrafter and DAV predict infinity values for distant buildings, resulting in missing structures, whereas our method consistently produces regular structures and plausible depth values, even when the ground truth exceeds the LiDAR sensor's range.

\subsection{Ablation Study}

\begin{table}
\centering
\caption{
  \textbf{Ablation study on the effectiveness of point map VAE.}
  }
\label{tab:ablate_depth}
\vspace{-0.5em}
\setlength{\tabcolsep}{2pt}
\begin{adjustbox}{width=\linewidth}
\begin{tabular}{l|cc|cc|cc|cc|cc}
\toprule 
\multirow{2}{*}{} & \multicolumn{2}{c|}{GMU Kitchen} & \multicolumn{2}{c|}{Monkaa} & \multicolumn{2}{c|}{Sintel} & \multicolumn{2}{c|}{Scannet} & \multicolumn{2}{c}{KITTI} \\
& $\text{Rel}^{d}\!\downarrow$& $\delta^{d}\!\uparrow$ & $\text{Rel}^{d}\!\downarrow$ & $\delta^{d}\!\uparrow$ & $\text{Rel}^{d}\!\downarrow$ & $\delta^{d}\!\uparrow$ & $\text{Rel}^{d}\!\downarrow$ & $\delta^{d}\!\uparrow$ & $\text{Rel}^{d}\!\downarrow$ & $\delta^{d}\!\uparrow$ \\
\midrule
w/o & 9.50 & 90.8 & 16.1 & 79.2 & 25.2 & 72.8 & 8.11 & 95.0 & \first{5.00} & 97.9 \\
w/ & \first{8.03} & \first{94.0} & \first{13.0} & \first{80.5} & \first{16.9} & \first{73.2} & \first{7.57} & \first{95.9} & 5.25 & \first{98.3} \\
\bottomrule
\end{tabular} 
\end{adjustbox}
  \vspace{-1.5em}
\end{table}

\begin{table*}[htbp]
\centering
\caption{
  \textbf{VAE reconstruction performance with different components.}
  Light gray background highlights our final VAE configuration.
  }
\label{tab:ablate_vae}
\vspace{-0.75em}
\begin{adjustbox}{width=\textwidth}
\begin{tabular}{cccc|cccc|cccc|cccc}
\hline 
\multicolumn{4}{c|}{ } & \multicolumn{4}{c|}{Scannet} & \multicolumn{4}{c|}{Sintel} & \multicolumn{4}{c}{Monkaa}   \\
Representation & $\mathcal{L}_\text{ms}$ & Temporal layers & Latent alignment & $\text{Rel}^{p}\!\downarrow$& $\delta^{p}\!\uparrow$ & $\text{Rel}^{d}\!\downarrow$& $\delta^{d}\!\uparrow$ & $\text{Rel}^{p}\!\downarrow$ & $\delta^{p}\!\uparrow$ & $\text{Rel}^{d}\!\downarrow$& $\delta^{d}\!\uparrow$ & $\text{Rel}^{p}\!\downarrow$& $\delta^{p}\!\uparrow$ & $\text{Rel}^{d}\!\downarrow$& $\delta^{d}\!\uparrow$\\
\hline 
\cref{eq:raw_pmap} & \checkmark &  &\checkmark &  7.67 & 99.8 & 2.03 & 99.8 &  8.25 & 94.4 & 6.46 & 94.8 & 6.05 & \second{99.5} & 3.93 & \second{99.4}\\

\cref{eq:pmap} &\checkmark  & & \checkmark &  \first{1.63} & 99.8 & \second{1.51} & 99.7 & \second{4.24} & 97.8 & 4.02 & \second{97.9} & \second{2.19} & \first{99.6} & \second{2.07} & \first{99.5}\\

\cref{eq:pmap} & & & \checkmark &  2.95 & 99.8 & 2.31 & 99.6 & 5.32 & 97.3 & 4.72 & 97.5 & 2.83 & \second{99.5} & 2.68 & \second{99.4}\\

\cellcolor{light}\cref{eq:pmap} & \cellcolor{light}\checkmark & \cellcolor{light}\checkmark & \cellcolor{light}\checkmark & \cellcolor{light}\second{1.65} & \cellcolor{light}\first{99.9} & \cellcolor{light}\first{1.47} & \cellcolor{light}\first{99.9} & \cellcolor{light}\first{3.45} & \cellcolor{light}\first{98.1} & \cellcolor{light}\first{3.06} & \cellcolor{light}\first{98.1} & \cellcolor{light}\first{2.01} & \cellcolor{light}\second{99.5} & \cellcolor{light}\first{1.84} & \cellcolor{light}\second{99.4} \\


\cref{eq:pmap} & \checkmark & \checkmark &  &  1.95 & \first{99.9} & 1.77 & \first{99.9} & {4.48} & \second{98.0} & \second{3.78} & \second{97.9} & {2.94} & {99.1} & {2.64} & {98.9} \\

\hline
\end{tabular} 
\end{adjustbox}
\vspace{-0.5em}
\end{table*}

\begin{table*}[htbp]
  \centering
  \caption{
    \textbf{UNet prediction performance with different components.}
    Light gray background highlights our final UNet configuration.
    }
  \label{tab:ablate_unet}
  \vspace{-0.75em}
  \begin{adjustbox}{width=\textwidth}
  \begin{tabular}{cc|cccc|cccc|cccc}
  \hline 
  \multicolumn{2}{c|}{ } & \multicolumn{4}{c|}{GMU Kitchen} & \multicolumn{4}{c|}{Sintel} & \multicolumn{4}{c}{DDAD} \\
  Latent alignment & Per-frame geometry prior &$\text{Rel}^{p}\!\downarrow$& $\delta^{p}\!\uparrow$ & $\text{Rel}^{d}\!\downarrow$ & $\delta^{d}\!\uparrow$ & $\text{Rel}^{p}\!\downarrow$ & $\delta^{p}\!\uparrow$ & $\text{Rel}^{d}\!\downarrow$ & $\delta^{d}\!\uparrow$ & $\text{Rel}^{p}\!\downarrow$ & $\delta^{p}\!\uparrow$ & $\text{Rel}^{d}\!\downarrow$ & $\delta^{d}\!\uparrow$ \\
  \hline
   & \checkmark  & 9.51 & 93.6 & 9.26 & 92.6 & \first{25.4} & \first{65.5} & 23.0 & 61.7 & \first{14.3} & 90.1 & 12.5 & 89.6\\
  \checkmark &  & 12.9 & 88.7 & 12.0 & 84.7 & 34.4 & 38.8 & 24.7 & 57.9 & 25.5 & 58.7 & 15.7 & 78.3 \\
  
  \cellcolor{light}\checkmark &  \cellcolor{light}\checkmark & \cellcolor{light}\first{8.52} & \cellcolor{light}\first{94.3} & \cellcolor{light}\first{8.30} & \cellcolor{light}\first{93.6} & \cellcolor{light}{25.6} & \cellcolor{light}{64.9} & \cellcolor{light}\first{22.6} & \cellcolor{light}\first{63.7} & \cellcolor{light}{15.0} & \cellcolor{light}\first{90.6} & \cellcolor{light}\first{12.0} & \cellcolor{light}\first{90.4} \\

  \hline \hline 
  \multicolumn{2}{c|}{DUSt3R} & 22.2 & 68.2 & 21.6 & 64.0 & 43.9 & 35.6 & 41.3 & 36.0 & 37.8 & 37.3 & 32.3 & 46.5\\
  \multicolumn{2}{c|}{Ours(G) + DUSt3R} & \first{12.2} & \first{90.4} & \first{11.5} & \first{88.1} & \first{34.4} & \first{39.9} & \first{26.2} & \first{55.8} & \first{28.7} & \first{48.8} & \first{15.9} & \first{80.9}\\
  \hline
  \end{tabular}
  \end{adjustbox}
  
  \vspace{-1.2em}
  \end{table*}


\noindent\textbf{Effectiveness of point map VAE.}
We conduct ablation studies by removing the point map VAE and using SVD's VAE to encode normalized disparity maps while keeping other components unchanged and retraining the model.
As shown in~\cref{tab:ablate_depth}, the estimated depth maps exhibit a significant performance drop across various datasets, except for KITTI, where the ground truth is constrained by the LiDAR sensor's range.
The visual comparison in~\cref{fig:ablate_depth} reveals that the performance decline is due to information loss from compressing unbounded depth values into a bounded range, leading to the neglect of distant objects.



\begin{figure}[t]
  \centering
  \includegraphics[width=0.99\linewidth]{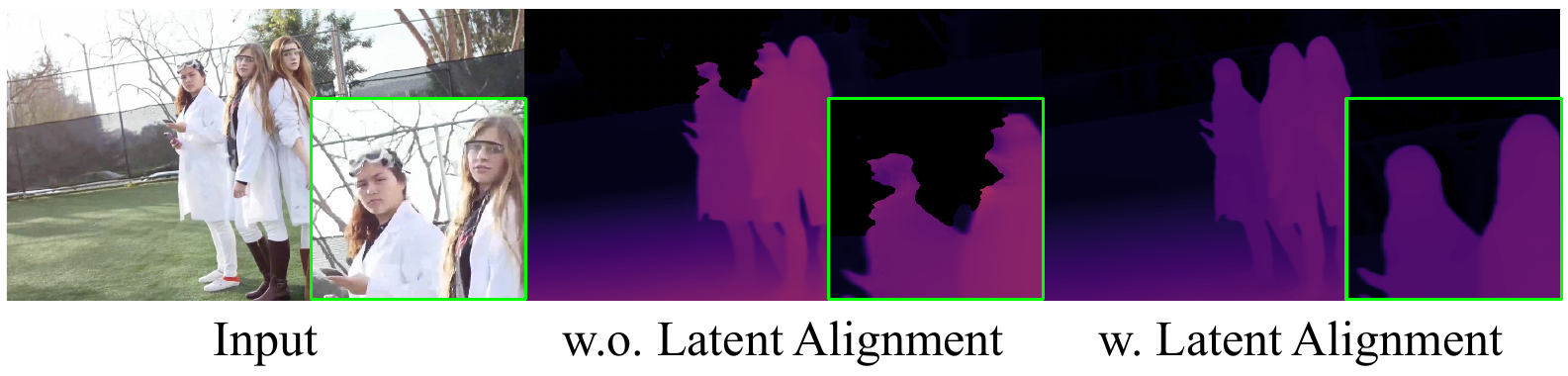}
  \vspace{-0.5em}
   \caption{
    \textbf{Effectiveness of latent alignment.}
    }
   \label{fig:ablate_la}
   \vspace{-1.5em}
\end{figure}

\noindent\textbf{Components in point map VAE.}
We perform ablation studies to examine the effectiveness of the point map representation, multi-scale loss $\mathcal{L}_\text{ms}$, temporal layers in the decoder, and latent alignment.
As shown in~\cref{tab:ablate_vae}, the decoupled point map representation~\cref{eq:pmap} markedly enhances reconstruction fidelity.
Results in the second to fourth rows also highlight the importance of multi-scale supervision in the spatial domain and contextual information in the temporal domain.
Eliminating the latent alignment component not only increases point map errors (see the last two rows of~\cref{tab:ablate_vae}), but also hinders effectively leveraging video diffusion priors.
As shown in~\cref{tab:ablate_unet} and~\cref{fig:ablate_la}, latent alignment substantially improves the quality and robustness of point map predictions.

\noindent\textbf{UNet design.}
We investigate the impact and robustness of per-frame geometry priors derived from MoGe~\cite{wang2024moge} by excluding them from the UNet input and replacing MoGe with DUSt3R~\cite{wang2024dust3r}.
As shown in~\cref{tab:ablate_unet}, per-frame priors benefit the model across diverse scenarios by compensating for limited camera intrinsics in the training data.
Moreover, replacing MoGe with DUSt3R also consistently improves performance, confirming the robustness of our method to different priors.
Besides, we present two variants of the UNet: one with the diffusion framework (noted as Ours(G)) and the other with a deterministic scheme (noted as Ours(D)).
As shown in~\cref{tab:comp_point} and~\cref{tab:comp_depth}, the deterministic approach exhibits slightly lower accuracy but achieves a $1.1\times$ acceleration in inference speed, \eg 4.1 v.s. 3.7 FPS at a $448\times768$ resolution on our experimental setup.
Users may choose one of the two variants based on their requirements for speed or accuracy.

%% file: sections/application.tex




\subsection{Applications}

\noindent\textbf{3D/4D reconstruction.}
With our temporally consistent, high-quality point maps, we enable 3D/4D reconstruction, whose cornerstone is the camera pose estimation. 
To this end, if dynamic objects exist, we first obtain their masks using SegmentAnything~\cite{kirillov2023segment} and XMem~\cite{cheng2022xmem}.
Then, we detect interest points in the static regions with SuperPoint~\cite{detone2018superpoint} and track them via SpaTracker~\cite{xiao2024spatialtracker}.
Finally, we optimize the camera poses with the established correspondences by 3D geometric constraints, taking only a few minutes to converge.
Examples of 3D/4D reconstruction are shown in~\cref{fig:teaser} and supplementary materials.

\begin{figure}[t]
  \centering
  \includegraphics[width=0.99\linewidth]{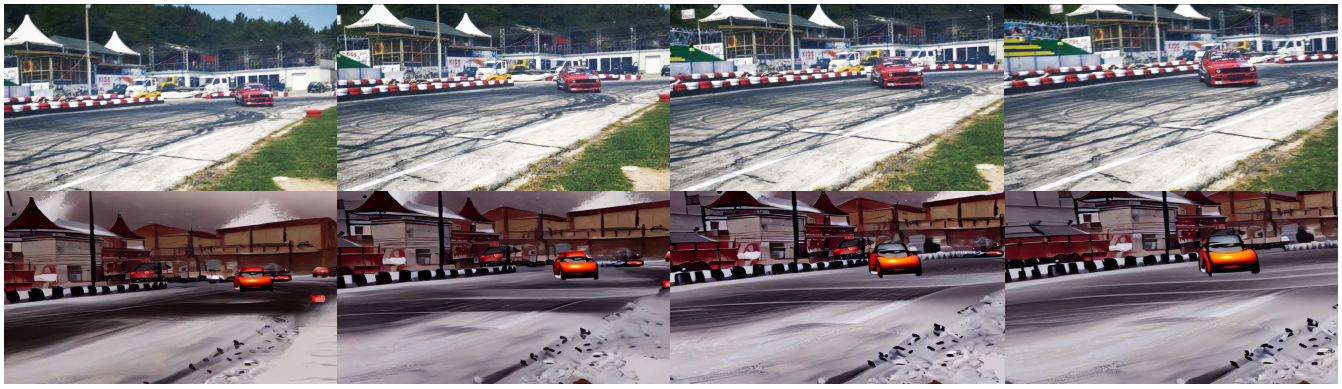}
    \vspace{-0.5em}
   \caption{
    \textbf{Application of depth-conditioned video generation.}
    The prompt is ``a car is drifting on roads, snowy day, artstation".
   }
   \label{fig:application}
   \vspace{-1em}
\end{figure}

\noindent\textbf{Depth-conditioned video generation.}
Depth sequences are pivotal to controllable video generation, capturing the inherent 3D structures of videos.
Our consistent depth maps serve directly as conditioning inputs in existing depth-driven methods (\eg, Control-A-Video~\cite{chen2023controlavideo}), enabling creative outputs as shown in~\cref{fig:application}.

%% file: sections/conclusion.tex
\section{Conclusion}


We present {\ourMethod}, a novel method that estimates temporally consistent, high-quality point maps from open-world videos, facilitating downstream applications such as 3D/4D reconstruction and depth-based video editing or generation. 
Our core design is a point map VAE that learns a latent space agnostic to original video latent distribution, enabling effective encoding and decoding of unbounded point map values.
We also introduce a decoupled point map representation to eliminate the location-dependent characteristics of point maps, enhancing the robustness to resolutions and aspect ratios.
Furthermore, we integrate a per-frame geometry prior conditioned diffusion model to model the distribution of point sequences conditioned on the input videos.
Comprehensive evaluations confirm that our method outperforms prior methods in performance and generalization.
Its main limitation is relatively high computational and memory overhead due to the large model size.

%% file: supps/dataset.tex
\section{Datasets}

\subsection{Training Datasets}

\begin{table}
\setlength{\tabcolsep}{2pt}
\begin{center}
\caption{\textbf{An overview of the training datasets.}}
\label{tab:train_dataset}
\resizebox{1.0\linewidth}{!}{
\begin{tabular}{cccc}
\hline 
Dataset & Domain & \#Frames & \#Videos\\
\hline
3DKenBurns~\cite{niklaus20193d} & In-the-wild & 76K & 526 \\
DynamicReplica~\cite{karaev2023dynamicstereo} & Indoor/Outdoor & 145K & 1126 \\
GTA-SfM~\cite{wang2020gtasfm} & Outdoor/In-the-wild & 19K & 234 \\
Hypersim~\cite{roberts2021hypersim} & Indoor & 75K & $\times$ \\
IRS~\cite{wang2021irs} & Indoor & 103K & 722 \\
MatrixCity~\cite{li2023matrixcity} & Outdoor/Driving & 452K & 3029 \\ 
MidAir~\cite{fonder2019mid} & Outdoor/In-the-wild & 357K & 2433 \\
MVS-Synth~\cite{huang2018deepmvs} & Outdoor/Driving & 12K & 120 \\
Spring~\cite{mehl2023spring} & In-the-wild & 5K & 49 \\
Structured3D~\cite{zheng2020structured3d} & Indoor & 71K & $\times$ \\
Synthia~\cite{ros2016synthia} &  Outdoor/Driving & 178K & 1276 \\
TartanAir~\cite{wang2020tartanair} & In-the-wild & 306K & 2245 \\
UrbanSyn~\cite{gomez2023all} & Outdoor/Driving & 7K & $\times$ \\
VirtualKitti2~\cite{cabon2020virtual} & Driving & 43K & 320 \\ 
\hline
Total & - & 1.85M & 12K \\
\hline
\end{tabular} }
\end{center} 
\vspace{-0.3cm}
\end{table}

We collect 14 open-source synthetic RGBD datasets to facilitate the training of {\ourMethod}, among which 11 can be composited into video sequences. To construct the training video dataset, we extract non-overlapping segments with a sequence length not exceeding 150 frames. An overview of the training datasets is provided in ~\cref{tab:train_dataset}, categorized into four distinct domains: indoor, outdoor, in-the-wild and driving scenarios. It is noteworthy that the frame count may slightly differ from the original datasets, owing to the exclusion of invalid frames. To ensure computational efficiency and adhere to GPU memory constraints, we preprocess all images and videos to a standardized resolution of $320 \times 640$. Specifically, we apply cover resizing while preserving the original aspect ratio, followed by center cropping to achieve the desired resolution. Additionally, we implement random resizing as a technique for augmenting camera intrinsics.

\subsection{Evaluation Datasets}

We exhaustively evaluate {\ourMethod} and previous state-of-the-art methods using seven datasets with ground truth labels that remain entirely unseen during the training phase. Notably, to ensure compatibility with most baselines, such as MoGe~\cite{wang2024moge} and UniDepth~\cite{piccinelli2024unidepth}, which necessitate an input image aspect ratio of less than 2, we preprocess the evaluation datasets in the following manner: 
\begin{itemize}
    \item \textbf{GMU Kitchens}~\cite{georgakis2016multiview}: All scenarios are employed for evaluation. For each scenario, we extract 110 frames with a stride of 2 to ensure extensive spatial coverage while preserving temporal coherence. To reduce memory usage during evaluation, we downsample the generated 1920p videos and ground truth depth maps to a resolution of $960\times540$.
    \item \textbf{ScanNet}~\cite{dai2017scannet}: Following DepthCrafter~\cite{hu2024-DepthCrafter}, we select 100 scenes from the test split for evaluation, wherein each video comprises 90 frames with a frame stride of 3. Due to the discrepancy in resolutions between the RGB images and depth maps, we first resize the RGB images to align with the depth maps, followed by center cropping to remove the black space around RGB images, yielding videos of resolution $624\times464$ .
    \item \textbf{DDAD}~\cite{ddad2020}: All 50 sequences from the validation split of the DDAD dataset are utilized for evaluation, with sequence lengths of either 50 or 100 frames. Owing to the high memory demands of the raw resolution $1936\times1216$, we apply center cropping to reduce the resolution to $1920\times1152$, followed by downsampling to $640\times384$ for evaluation. The ground truth depth maps, acquired via LiDAR sensors, are inherently sparse; consequently, the preprocessing has negligible influence on the comparative analysis of various methods.
    \item \textbf{KITTI}~\cite{Geiger2013IJRR}: All sequence in the valid split of depth annotated dataset are used evaluation. For excessively long video sequences, we extract the initial 110 frames, resulting in 13 videos with sequence lengths ranging between 67 and 110 frames. Given that the original resolution of $1242\times375$ fails to conform to the aspect ratio requirements of most baseline methods, we apply center cropping to achieve a resolution of $736\times368$.
    \item \textbf{Monkaa}~\cite{mayer2016large}:We select 9 scenes from the original dataset for evaluation, truncating each video sequence to 110 frames while maintaining the original resolution of $960\times540$. To derive valid masks, we manually annotate the sky regions within each sequence.
    \item \textbf{Sintel}~\cite{sintel}: All sequences within the training split are employed for evaluation, with sequence lengths ranging between 21 and 50 frames. Given the original resolution of $1024 \times 436$ for each image, we apply cropping to achieve a standardized resolution of $872 \times 436$.
    \item \textbf{DIODE}~\cite{diode_dataset}: We utilize all 771 images from the validation split of DIODE for evaluation purposes. To address the noisy values along the edges of objects within the depth maps, we employ a Canny filter to detect edge regions, subsequently refining the valid masks based on the filtering outcomes.
    
\end{itemize}

%% file: supps/losses.tex
\section{Loss Functions of VAE and UNet}

To train the point map VAE, we define the loss function $\mathcal{L}_\text{pmap}$ to measure the reconstruction errors of point maps. The reconstruction loss $\mathcal{L}_\text{recon}$ for each valid pixel is defined as the $L_1$ norm
\begin{align}    
    \mathcal{L}_\text{recon} = \sum_{p\in \mathcal{M}} ||z_p - \widehat{z}_p||_1 + \sum_{p\in \mathcal{M}} ||\theta_\text{diag} - \widehat{\theta}_\text{diag}||_1
\end{align}
where $\mathcal{M} = \{p|\mathbf{m}(p) = 1\}$ and $\widehat{z}_{p}, \widehat{\theta}_\text{diag}$ are the reconstructed values at pixel $p$. To enhance surface quality, we additionally supervise the normal maps derived from the reconstructed point maps and the ground truth:
\begin{align}
\mathcal{L}_\text{n} = \sum_{p\in \mathcal{M}} (1-n_p \cdot \widehat{n}_{p})
\end{align} 
To enhance supervision for local geometry, we draw inspiration from MoGe~\cite{wang2024moge} and propose a multi-scale depth loss function that measures the alignment between reconstructed and ground truth depth maps within local regions $\mathcal{H}_\alpha$, parameterized by scale $\alpha$
\begin{align}
    \mathcal{L}_\text{ms} = \sum_{\mathcal{H}_\alpha}\sum_{p\in \mathcal{H}_\alpha \& p\in \mathcal{M}} ||(z_{p} - \overline z_{p,\mathcal{H}_\alpha}) - (\widehat{z}_{p}-\widetilde{z}_{p,\mathcal{H}_\alpha})||_1
    \label{eq:patch_loss}
\end{align}
Here, $\overline z_{p,\mathcal{H}_\alpha}$ and $\widetilde{z}_{p,\mathcal{H}_\alpha}$ are the mean value of predicted and ground truth depth map defined on local region $\mathcal{H}_\alpha$. In practice, we split video frames into non-overlapped patches of size $\frac{W}{\alpha}\times \frac{H}{\alpha}$ to define the local regions. The reconstruction objective $\mathcal{L}_\text{pmap}$ is thus given by 
\begin{equation}
    \mathcal{L}_\text{pmap} = \mathcal{L}_\text{recon} + \mathcal{L}_\text{ms} + \lambda_\text{n}\mathcal{L}_\text{n}
\end{equation}
Following LayerDiffuse~\cite{zhang2024layerdiffuse}, we apply the frozen decoder $\mathcal{D}_\text{SVD}$ to measure the extent to which the latent offset disrupts the modified latent distribution during training, given by
\begin{equation}
\mathcal{L}_\text{identity} = ||\widetilde{\mathbf x}_\text{disp}-\widehat{\mathbf x}_\text{disp}||_2^2 = ||\widetilde{\mathbf x}_\text{disp} - \mathcal D_\text{SVD}(\mathbf z_\text{pmap})||_2^2    
\label{eq:iden_reg}
\end{equation}
where $||\cdot||_2^2$ denotes the mean square loss function. Additionally, we introduce a mask loss to regularize the reconstructed valid mask:
\begin{equation}
    \mathcal{L}_\text{mask} = ||\widehat{\mathbf{m}} - \mathbf{m}||_2^2
\end{equation}
where $\mathbf{m}\in \mathbb{R}^{T\times H \times W}$ is the ground truth valid mask. The final training objective of VAE is defined as
\begin{equation}
    \mathcal{L}_\text{VAE} = \mathcal{L}_\text{identity} + \mathcal{L}_\text{pmap} + \lambda_\text{mask}\mathcal{L}_\text{mask}
\end{equation}

To finetune the UNet $D_\theta$ parameters on the adjusted latent space obtained by our proposed point map VAE, we employ the objective $\mathcal{L}_\text{UNet}$ for supervision, written as
\begin{small}
\begin{multline}
    \mathbb{E}_{\mathbf{z}_t\sim p(\mathbf{z},\sigma_t), \sigma_t\sim p(\sigma)}[\lambda_{\sigma_t}||D_\theta(\mathbf{z}_t; \sigma_t, \mathbf{z}_\mathbf{v}, \mathbf{z}_\text{prior})-\mathbf{z}_\text{pmap}||_2^2]
    \label{eq:unet_loss}
\end{multline}
\end{small}

\noindent Here the noisy latent input $\mathbf{z}_{t}$ is generated by adding Gaussian noise $n$ to the latent code $\mathbf{z}_\text{pmap}$. $\mathbf{z}_\mathbf{v}$ is the conditional latent code of input video. $\mathbf{z}_\text{prior}$ denotes the per-frame geometry priors provided by MoGe~\cite{wang2024moge}. $\sigma_t$ denotes noise level at time $t$, satisfying $\log \sigma_t \sim \mathcal{N}(P_\text{mean},P_\text{std})$ with $P_\text{mean}=0.7$ and $P_\text{std} = 1.6$ adopted in the EDM~\cite{karras2022edm} noise schedule and $\lambda_{\sigma_t}$ is a weight parameter at time $t$. 


%% file: supps/implementation.tex
\section{More Implementation Details}

For the point map VAE design, we reuse the architecture of SVD's VAE with minor modification: we adopt zero convolution initialization~\cite{zhang2023adding} to the output convolution layer of encoder and apply a scale factor of 0.1 to ensure that latent offsets do not disrupt the latent distribution during the initial stage of training. Inspired by the training strategy of SVD, we first train the model from scratch with an AdamW~\cite{loshchilov2017decoupled} optimizer on RGBD images, with a fixed learning rate of 1e-4 for 40K iterations. Then, we finetune the temporal layers in the decoder for another 20K iterations on video data. The batch sizes are set to be 64 and 8 for the respective stages, with sequence lengths randomly sampled from $[1, 8]$ for video data in the second stage. For the UNet denoiser, we initialize UNet with the pretrained parameters provided by DepthCrafter~\cite{hu2024-DepthCrafter}, finetuning it with a learning rate of 1e-5 and a batch size of 8. We train our diffusion UNet in two stages, where we first train it on videos with sequence lengths sampled from [1, 25] frames to adapt the model to our generation task, and then solely finetune the temporal layers with the sequence length randomly sampled from [1, 110] frames due to the limitation of GPU memory. After training, the UNet can process videos with varying lengths (e.g., 1 to 110 frames) at a time. Both components are trained on $320 \times 640$ images or videos for efficiency, with random resizing and center cropping applied for data augmentation and resolution alignment. All trainings are conducted on 8 GPUs, with the entire process requiring about 3 days.

%% file: supps/camera_pose.tex
\section{Camera Pose Estimation}

To recover camera poses from point maps, we need to establish correspondences of the static background across frames. We first obtain the dynamic object masks by annotating the first frame using SegmentAnything~\cite{kirillov2023segment}, and then apply XMem~\cite{cheng2022xmem}, a robust method for video object segmentation, to generate the dynamic target masks for the subsequent frames. Given the dynamic masks, we adopt SuperPoint~\cite{detone2018superpoint} to detect reliable points of interest in the first frame and filter out those points that belong to the dynamic objects. After that, we employ SpaTracker~\cite{xiao2024spatialtracker} to generate the 2D trajectory of each point, which is subsequently used to form the constraints for the camera pose optimization. Let $p_t$ denote the XY coordinate of a 2D trajectory at time step $t$, the 2D point $p_t$ can be lifted to the world coordinate $\widetilde p_t$ using the following transformation
\begin{equation}
    \widetilde p_t = W_t^{-1} \pi_{K_t}^{-1}(p_t, D(p_t))
\end{equation}
Here $W_t$ denotes the camera pose at time step $t$, $D(\cdot)$ denotes the scale-invariant depth value obtained from our predicted point maps and $\pi_{K_t}^{-1}$ refers to the back-projection of the 2D point to camera coordinate with camera intrinsic $K$, which can also be estimated from the point maps. For time step $t'$, the 2D projected coordinate should align with the trajectory position at timestep $t'$. Therefore, we formulate the camera pose estimation as the following problem
\begin{small}
\begin{equation}
    \min_{W_1...W_T} \sum_{i,j\in [1...T]}||\pi_{K_j}W_{j}W_i^{-1} \pi_{K_i}^{-1}[p_i, D(p_i)]- [p_{j}, D(p_j)]||_2^2
\end{equation}
\end{small}

\noindent Due to the sequence length limitation of SpaTracker (12 for each segment), we apply a shifted window strategy with 6 overlapping frames to regularize the optimization of all camera poses. The optimization process for each scene takes from less than 1 minute to several minutes, relying on the number of frames.

%% file: supps/limitations.tex
\section{Limitations}

\begin{table}
\setlength{\tabcolsep}{2pt}
\begin{center}
\caption{\textbf{Inference time.of different components on $448\times768$ videos with 110 frames.}}
\label{tab:infer}
\resizebox{1.0\linewidth}{!}{
\begin{tabular}{c|cccc|c}
\hline 
Method & Per-frame Prior & Encoder & UNet & Decoder & Total \\
\hline
Ours(G) & 0.1 & 0.04 & 0.04 & 0.08 & 0.27s/frame \\
Ours(D) & 0.1 & 0.04 & 0.01 & 0.08 & 0.24s/frame \\
\hline
\end{tabular} }
\end{center} 
\vspace{-0.3cm}
\end{table}

The major limitation of our method is the expensive computation and memory cost, primarily attributing to the large model size inherent in both the VAE and U-Net architectures. As shown in ~\cref{tab:infer}, we provide a comparison of the inference times of different components in \ourMethod. Our experiments are conducted on a single GPU, revealing that the decoder of the point map VAE is the bottleneck during inference. How to design a lightweight decoder capable of producing temporally consistent outputs will be a focal point of our future works.

%% file: supps/results.tex
\section{More results}

\begin{figure*}[t]
  \centering
  \includegraphics[width=0.99\linewidth]{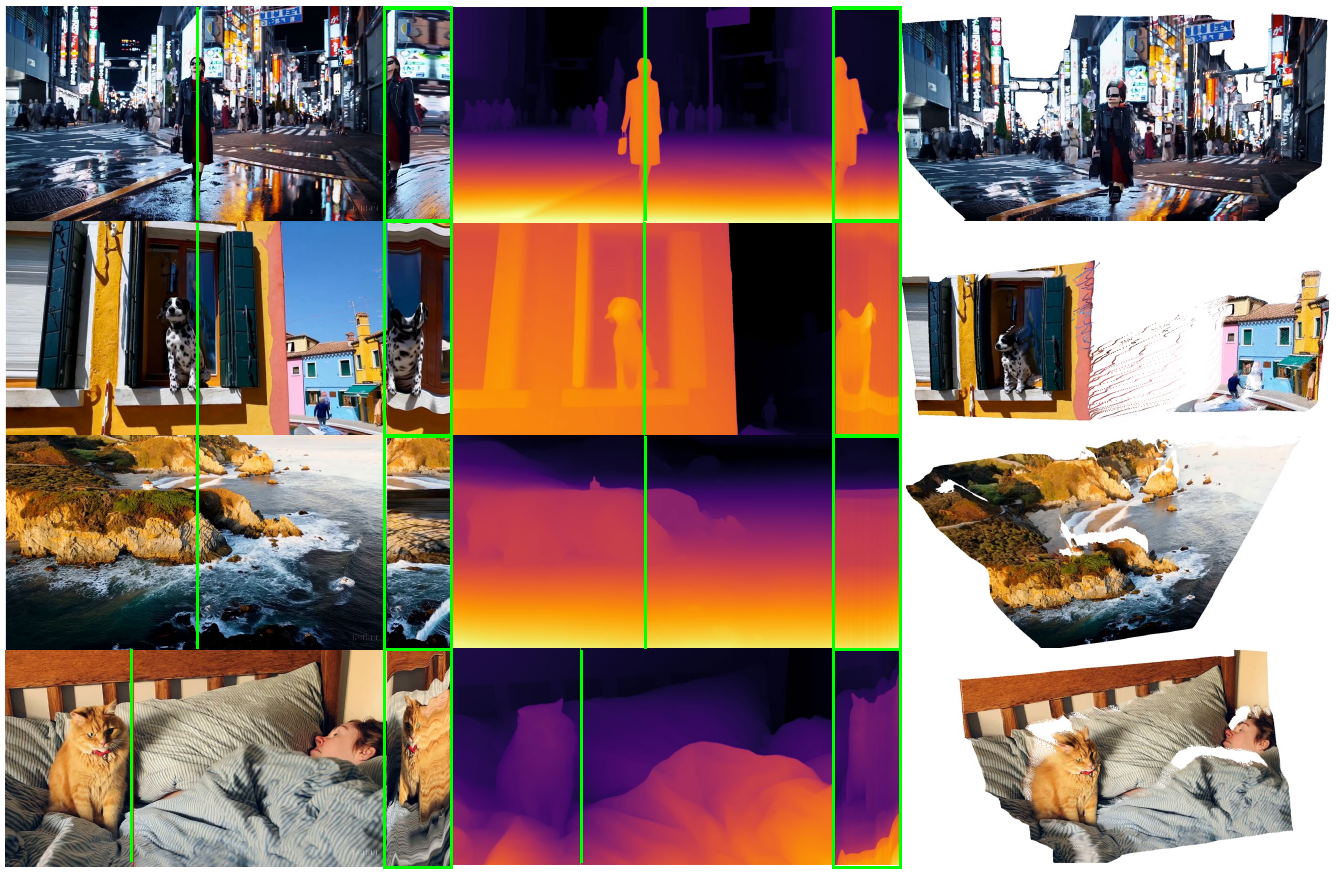}

   \caption{\textbf{Visual results on Sora-generated videos.}  The rows from left to right are the input videos, the disparity maps and the point cloud of the first frame.}
   \label{fig:sora}
\end{figure*}

\begin{figure*}[t]
  \centering
  \includegraphics[width=0.99\linewidth]{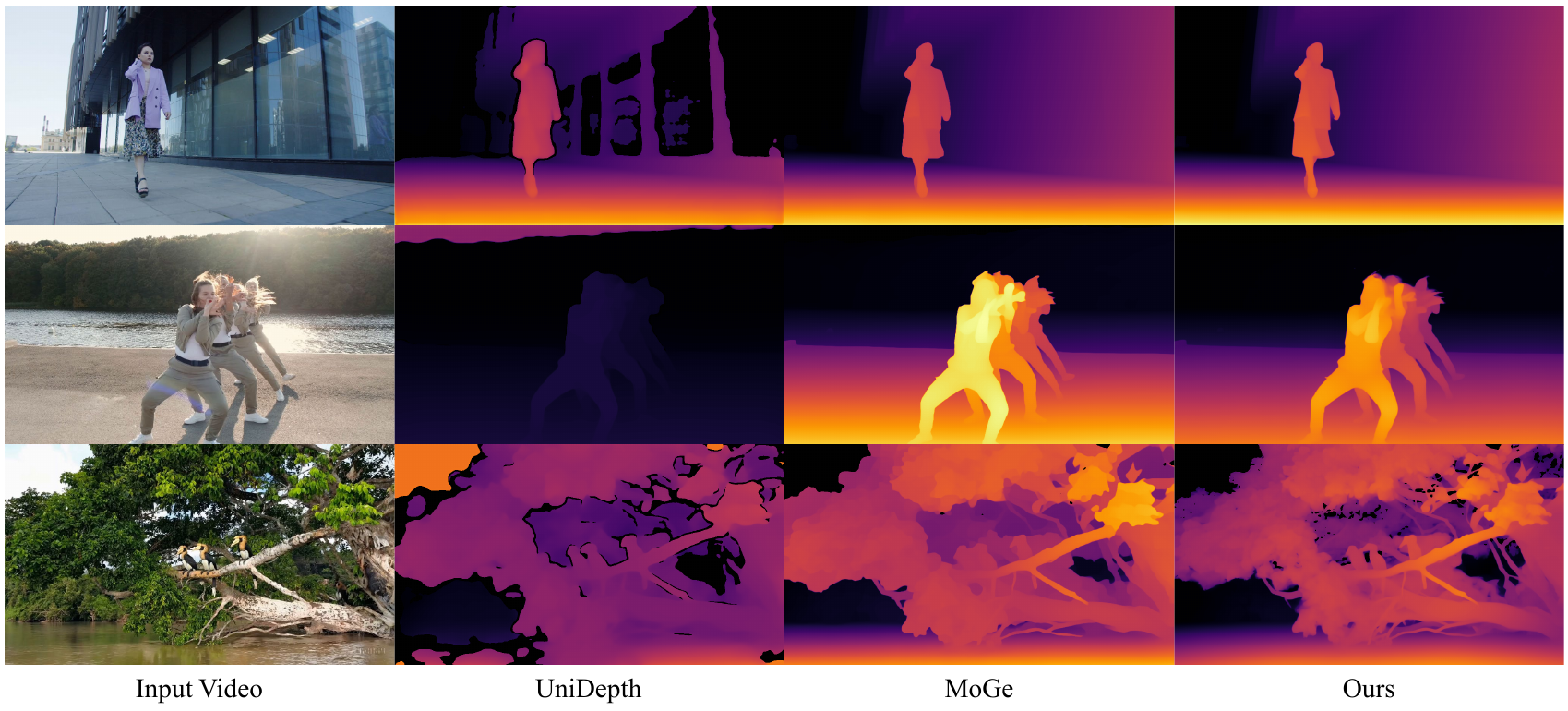}

   \caption{\textbf{Visual comparison with monocular geometry estimation methods.} All point maps are converted to disparity maps for better visualization the sharpness of depth prediction.}
   \label{fig:comp}
\end{figure*}

\begin{figure*}[t]
  \centering
  \includegraphics[width=0.99\linewidth]{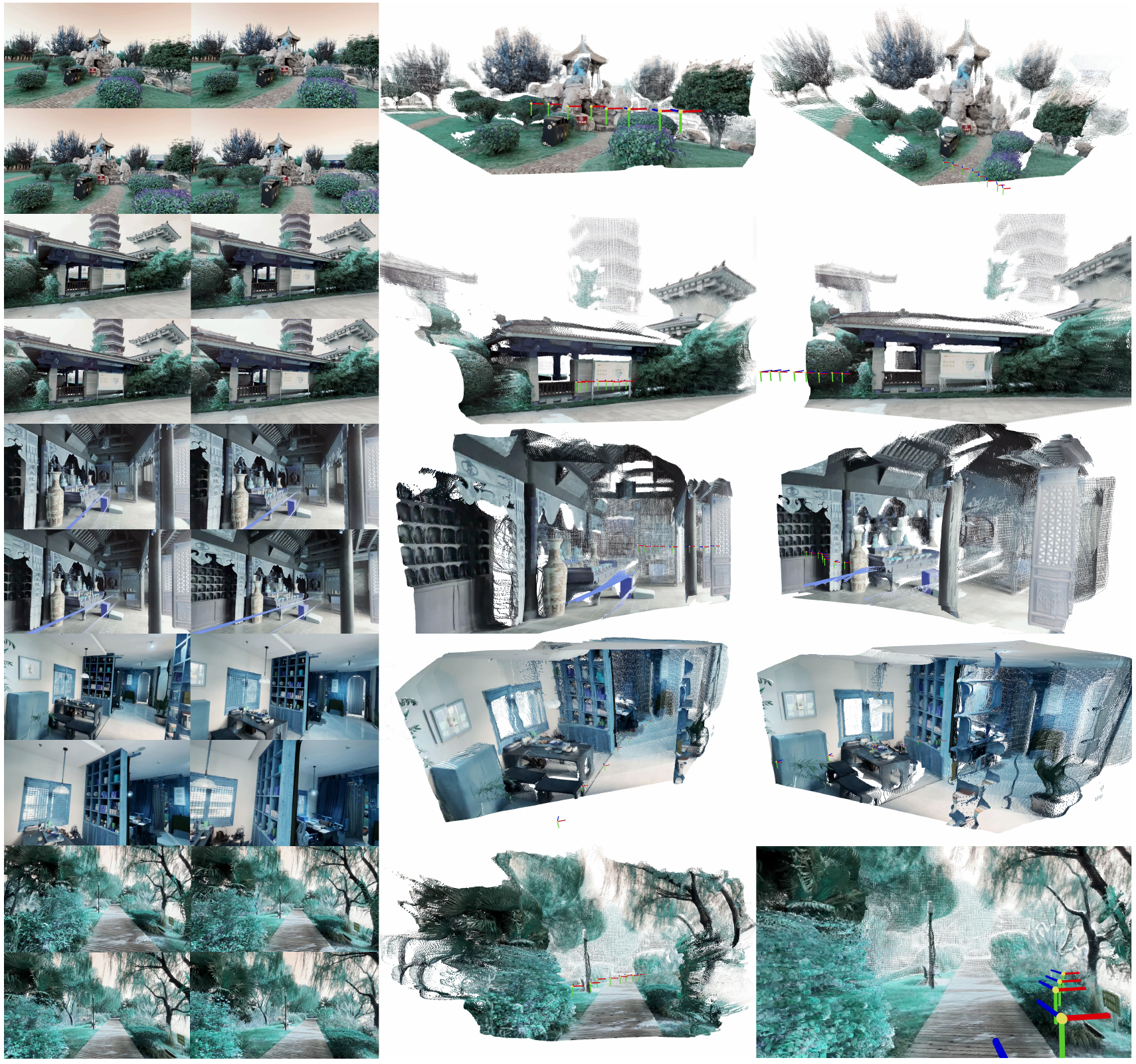}

   \caption{\textbf{Visual results on DL3DV~\cite{ling2024dl3dv} with camera poses estimated from the output point maps.} We concatenate 8 aligned point maps from the original point map sequence for visualization.}
   \label{fig:static}
\end{figure*}

\begin{figure*}[t]
  \centering
  \includegraphics[width=0.99\linewidth]{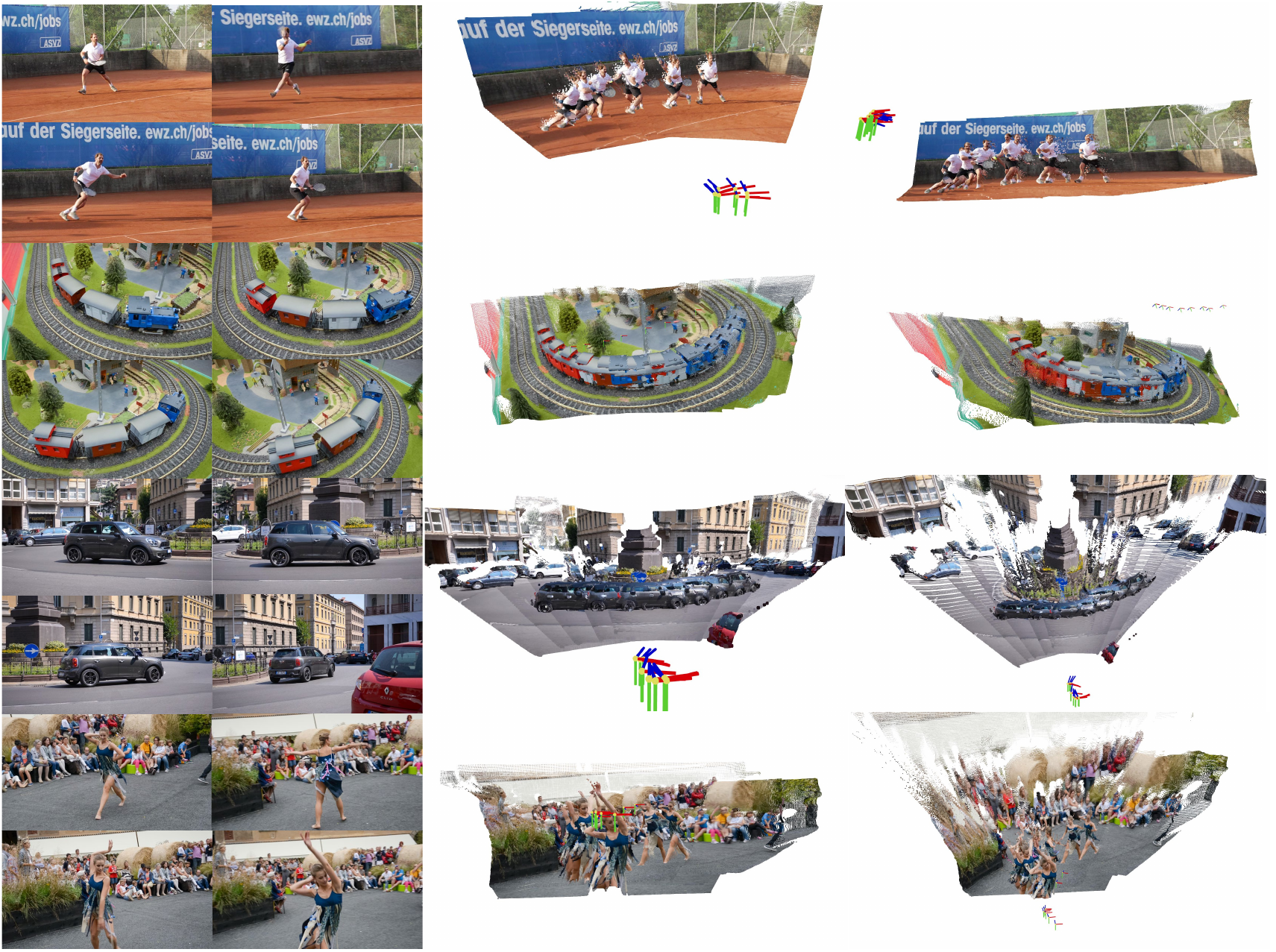}

   \caption{\textbf{Visual results on DAVIS~\cite{Perazzi2016davis} with camera poses estimated from the output point maps.} We concatenate 8 aligned point maps from the original point map sequence for visualization.}
   \label{fig:dynamic}
\end{figure*}

In the following pages, we provide more visual results of our method. We provide more results on Sora~\cite{sora}-generated videos to demonstrate the temporal consistency and geometry quality of our method, as shown in ~\cref{fig:sora}. For comprehensive comparison with MGE methods, we provide a visual analysis in ~\cref{fig:comp}. Our method achieves robust and sharp point map estimation compared to other methods. In contrast, UniDepth~\cite{piccinelli2024unidepth} fails to segment the sky region from the input frames, while MoGe~\cite{wang2024moge} struggles to handle fine-grained structure. ~\cref{fig:static} and ~\cref{fig:dynamic} show the point maps aligned with the optimized camera poses, where the rows from left to right are 4 input frames uniformly sampled from the whole video and two views of aligned point maps in the world coordinates. We only provide the results of concatenating 8 point maps sampled from the predicted point sequences for better visualization.

